\documentclass{ar2e}
\usepackage{natbib}
\bibpunct{(}{)}{,}{a}{}{,}

\def\lesssim{\mathrel{\hbox{\rlap{\hbox{\lower4pt\hbox{$\sim$}}}\hbox{$<$}}}}
\def\gtrsim{\mathrel{\hbox{\rlap{\hbox{\lower4pt\hbox{$\sim$}}}\hbox{$>$}}}}

\newcommand{\msol}{\mbox{M$_{\odot}$}}
\newcommand{\zsol}{\mbox{Z$_{\odot}$}}

\newcommand{\rsol}{\mbox{R$_{\odot}$}}

\newcommand{\kms}{\mbox{$\rm{km}\,s^{-1}$}}
\newcommand{\logl}{\mbox{$\log L/{\rm L_{\odot}}$}}
\newcommand{\hii}{\mbox{H\,{\sc ii}}}
\newcommand{\nick}{\mbox{$^{56}$Ni}}
\newcommand{\ergsec}{\mbox{ergs\,s$^{-1}$}}

\begin{document}
\input epsf.tex    
\input epsf.def   
\input psfig.sty

%
%

\jname{Annual Review of Astronomy and Astrophysics}
\jyear{2009}
\jvol{}
\ARinfo{1056-8700/97/0610-00}
\title{Progenitors of core-collapse supernovae}

\markboth{Supernova Progenitors}{Smartt}

\author{Stephen J. Smartt\thanks{This is a preprint. 
    Published article is ARAA, 2009, 47, 63. See link on http://star.pst.qub.ac.uk/$\sim$sjs }
\affiliation{Astrophysics Research Centre, School of Mathematics and Physics, Queen's University Belfast, N. Ireland, BT7 1NN, United Kingdom}}


\begin{abstract}

Knowledge of the progenitors of core-collapse supernovae is a
fundamental component in understanding the explosions.  The
recent progress in finding such stars is reviewed. The minimum
initial mass that can produce a supernova has converged to 
 $8\pm1$\msol, from direct detections of red supergiant
progenitors of II-P SNe and the most massive white dwarf progenitors, 
although this value is model dependent. 
It appears that most type Ibc
supernovae arise from moderate mass interacting binaries. The highly
energetic, broad-lined Ic supernovae are likely produced by massive,
Wolf-Rayet progenitors.  There is some evidence to suggest that the majority
of massive stars above  $\sim$20\msol\ may collapse quietly to
black-holes and that the explosions remain undetected.  The recent
discovery of a class of ultra-bright type II supernovae and the direct
detection of some progenitor stars bearing luminous blue variable
characteristics suggests some very massive stars do produce highly
energetic explosions. The physical mechanism is open to debate and 
these SNe pose a challenge to stellar evolutionary theory.

\end{abstract}

\maketitle

\newpage
\section{Introduction}

Stellar explosions have shaped the nature of the visible Universe.
The chemical elements heavier than boron were created in stars and
propelled through the galactic interstellar medium by virtue of the
enormous kinetic energies liberated during stellar deaths.  The most
massive stars are the primary drivers of galactic chemical evolution
with for example $\sim$0.4\msol\ of oxygen ejected by every 15\msol\ star
\citep{1996ApJ...460..408T}. 
Such
stars (with masses more than about 7-10\msol) have long been thought
to produce supernovae (SNe) when their evolutionary path ends with a
core of iron and further nuclear burning no longer provides thermal
pressure to support the star.  Given the astrophysical knowledge at
the time, \cite{1934PNAS...20..254B} made a great leap of faith in predicting
their newly named super-novae in external galaxies were the deaths of
massive stars that produced neutron stars and high energy cosmic
rays. This paradigm has stood for more than seventy years with great
efforts invested to understand supernovae and their remnants.  A major
goal has been theoretically predicting what type of stars can produce
iron or oxygen-magnesium-neon cores and collapse to give these
explosions
\citep[for example, a non-exhaustive list of recent work is:][]{2002RvMP...74.1015W,2003ApJ...591..288H,2004MNRAS.353...87E,2004A&A...425..649H}. 
Observationally testing these models with measurements of the physical
characteristics of the progenitor stars alongside the explosion
parameters can constrain the theory.

The mechanism of conversion of gravitational potential energy from the
collapsing 1.4\msol\ Fe core (with a radius similar to that of the
earth) into a shock induced explosion has been the subject of intense
theoretical activity in the modern computational era. 
The bounce from the imploding mantle rebounding off the nuclear
density proto-neutron star does not inject enough energy to produce a
shock with enough momentum to reach the surface
\citep{1986ARA&A..24..205W,2007PhR...442...38J}. At the extreme
temperatures and densities in the collapsing core, neutrinos of all
three flavours are created with a total luminosity of around
$3\times10^{53}$\,ergs. Deposition of a small fraction of their energy has been
proposed as the energy source to drive the explosion 
\citep{2007PhR...442...38J} and
recent work has advocated the idea of acoustic vibrations of the 
proto-neutron star \citep{2006ApJ...640..878B}. 
The discovery of neutrinos from SN1987A confirmed
the collapsing core idea in spectacular fashion
\citep{1987PhRvL..58.1490H}.

The community is patiently waiting for a Galactic
core-collapse event to test this physics with, presumably, a strong
neutrino and gravitational wave signal. 
The youngest SN remnant in the galaxy G1.9+0.3 is of order 150\,yrs
old \citep{2008MNRAS.387L..54G} 
and we may have a long wait for the next. Constraints on the models of 
stellar evolution, chemical element synthesis and explosion mechanisms
thus rely on the studies of SNe and their progenitor stars in 
other galaxies in the Local Universe. Supernovae from massive stars 
(CCSNe) have observed kinetic energies of typically $\sim10^{51}$\,ergs
and their integrated luminosities are usually 1-10\% of this value. 
However they display a huge range in their physical characteristics, 
including chemical composition of the ejected envelope, 
kinetic energy, radiated energy and the
explosively created radioactive composition 
($^{56}$Ni, $^{57}$Ni, $^{44}$Ti). Their 
properties are much more diverse than the thermonuclear type Ia 
SNe, which originate in white dwarf binary systems
\citep{2000ARA&A..38..191H}.  The 
energetically most extreme CCSNe are those associated with GRBs 
with kinetic energies of $2-5\times 10^{52}$\,ergs
\citep{2006ARA&A..44..507W}. 
A new class of 
ultra-bright SNe have total radiated energies $\sim10^{51}$\,ergs. 
(see Section\,\ref{subsect:LBV-SNe}). 

This diversity
reflects the large range of stellar types seen in the upper region 
of the Hertzsprung Russel Diagram (HRD) above $\sim$10\msol\
\citep{1994PASP..106.1025H,2003ARA&A..41...15M,2007ARA&A..45..177C}. 
Mass, binarity, metallicity, rotation rate, mass-loss rate and 
probably magnetic fields play critical roles in forming 
evolved objects of various radii, density profiles and surrounding
circumstellar medium 
\citep{1992ApJ...391..246P,2000ApJ...544.1016H,2004MNRAS.353...87E,2004A&A...425..649H,2005A&A...443..643Y}. 

The last decade has seen direct discoveries of many 
SN progenitors and an explosion in the numbers and diversity of SNe
discovered. This review will discuss the remarkable and rapid 
progress there has been in the last decade in 
identifying massive stars which have
subsequently exploded. For every 
nearby CCSNe which is discovered the global astronomical archives can be 
carefully searched to identify deep, high resolution images
of the CCSN position before explosion. Precise positioning of
the CCSN location on these pre-explosion images, with space
and ground-based large telescopes, offers the 
possibility of massive progenitor stars to be identified. 
Extraordinary theoretical progress has been made since Zwicky \& Baade
 by comparing stellar evolution models to 
lightcurve models of SN observations. Multi-wavelength 
surveys have discovered a huge diversity of explosions and outbursts. 
The possibility of glimpsing stars before they explode is a new and
powerful way to test theory.  
This review focuses on linking the knowledge we have gained from these observational discoveries to our knowledge of stellar evolution and the explosion parameters of SNe. It is a summary of the observational advances in the field, although some of the most interesting results come from interpretation of the observations using theoretical stellar evolution models. Where quantitative results depend on models, this is specifically mentioned. 

\section{Supernovae and resolved stellar populations in nearby galaxies}

\subsection{Supernova types and classification}

Supernovae are primarily classified by the appearance of their
optical spectra, usually around the time of peak brightness. A
thorough review of the types and the criteria used to classify them
is provided by \cite{1997ARA&A..35..309F}. The article points out that
the approach of is largely taxonomical and that there is value in
grouping similar SNe as variations of broad themes, rather than 
the introduction of new types. 
This has largely held true in the last ten years
and with many new observational discoveries the same SN types are by
and large used. The type I SNe are defined by the lack of hydrogen
features (either in emission or absorption). Type Ia SNe also show no
helium features but have a characteristic Si absorption
feature.  Type Ib have unambiguous signatures of helium and type Ic
SNe show no hydrogen or helium. Both Ib and Ic SNe show strong
features of the intermediate mass elements O, Mg and Ca. The type II
SNe are all defined by the presence of strong hydrogen lines and a
further sub-classification is made based on the lightcurves. Most type
II SNe can be further subdivided into the II-P SNe (which show a
plateau phase) and the type II-L which exhibit a linear decay after
peak brightness.  The type IIn SNe show hydrogen emission lines which
usually have multiple components of velocity and always have a strong
``narrow'' profile. There are often variations on these major
sub-categories, for example SN1987A is usually referred to as a
plateau-type event but was clearly peculiar. The type Ic SNe which are
associated with long gamma-ray bursts \citep{2006ARA&A..44..507W} all
show much broader lines than typical Ic SNe. They have been referred to 
as ``hypernova'' or broad-lined Ic SNe, due to them having large 
inferred kinetic energies. It can often be hard to distinguish between 
the Ib and Ic SNe and it is useful often to term the group Ibc SNe
and such terminology will be used in this review \citep{1997ARA&A..35..309F}. 
 Finally the IIb SNe
are those which begin with spectra like type II but evolve rapidly to 
exhibit He lines, 
and at the same time the H lines weaken and disappear. .

\subsection{Supernova surveys and explosion rates}
\label{subsect:rates}

The SNe for which one can directly attempt to identify progenitor stars
must be fairly nearby ($\lesssim$30\,Mpc) or the obvious problems of resolution and limiting
magnitude render searches meaningless. SN discoveries in catalogued galaxies 
in the local Universe (within about 140\,Mpc) have  
been dominated by the Lick Observatory Supernova Search over the past 
10 years \citep[LOSS;][]{2001ASPC..246..121F}, although a large number of well 
equipped and experienced (but unsalaried) astronomers with 0.3-0.7\,m telescopes
play a major role in 
discovering the closest explosions (e.g. K. Itagaki, T. Boles, T. Puckett
and R. Evans are amongst the most prodigious SN hunters working 
outside professional astronomical institutions). How many
nearby SNe are missed due to dust extinction in their hosts, or 
intrinsically faint luminosities, or neglected faint host galaxies,  
is still an open question. And how
those issues could affect the relative rates of different physical types
of explosion is also not well understood. This may be addressed in 
future all-sky imaging surveys with larger apertures such as 
Pan-STARRS and LSST \citep{2008arXiv0807.3070Y}. 

The existence of an initial mass function (IMF) with a slope that
strongly favours the formation of lower mass stars is now well
established to exist for massive stars in the Local Universe
\citep{2008arXiv0803.3154E,2003ARA&A..41...15M}. If CCSNe arise from
stars with masses greater than about 8\msol\ then the IMF necessitates
that stars in the 8-15\msol\ mass range should dominate the rate of 
explosions (60\%  of all, assuming a Salpeter slope of $\Gamma$=-1.35). 
Of course this is moderated by the effects of stellar evolution, binarity, 
initial rotation and metallicity. The frequency of 
occurrence of the different SN types and their true rate can give principal
constraints in establishing their nature. This section
will distinguish the measurement of {\em SN rates} (the true rate
of explosion per unit time and per unit of galaxy luminosity) and
the {\em relative frequency} of SN types
(the relative occurrence of each different subtype). Table\,\ref{table1}
lists the relative frequency of each sub-type from five different 
studies. 

The most reliable measurement of the local SN rate is still that of
\cite{1999A&A...351..459C}. They split the CCSN types into two broad
categories of type II and type Ibc and applied simple empirical bias
corrections to mitigate the effects of galaxy inclination and
extinction in their visual and photographic methods. Both
\cite{2007ApJ...661.1013L} and \cite{2005PASP..117..773V} have used
the discoveries of the LOSS only to estimate {\em relative frequencies} 
within distance
limits of about 30\,Mpc and 140\,Mpc (the limit for the LOSS)
respectively. They go further than \cite{1999A&A...351..459C} in
separating the IIn and IIb SNe from the overall type II
class. \cite{SECM08} have compiled all SNe discoveries in the
literature in a fixed 10.5 year period within galaxies with recessional
velocities $V_{\rm vir} <2000$\kms\ (corrected for Virgo infall, this implies a
distance of 28\,Mpc, assuming $H_{\rm0}=72$\,\kms) and reassessed 
all available data on the 92 CCNe to estimate the 
relative frequency of all the subtypes. The agreement between these four
studies, which have different distance and volume limits and sample a
wide range of SN surveys, is reasonably good and within the Poisson
statistical uncertainties there is no clear discrepancy between them.
\cite{2008ApJ...673..999P} caution that their sample of SNe in SDSS
star-forming galaxies would suggest that the ratio of the frequency of Ibc
to II ($N_{\rm Ibc}/N_{\rm II}$) goes down 
from $0.4\pm0.1$ at solar metallicity (\zsol)
to $0.1\pm0.1$ at a
metallicity of 0.3\zsol. The results in Table\,1 effectively average 
over metallicities between about
0.3-2\zsol\ (see Smartt et al. 2009 for a discussion). The agreement 
between the studies suggests that the  
relative frequencies (averaged over near solar metallicities) 
of the subtypes are now reliably determined.
In the future the challenge will be to determine metallicity
dependent rates with better measurement resolution, more statistics
and more accurate nebular oxygen abundances of the SNe environments.

An important question is how complete the local samples of SNe are.
At the distance limits of 28-30\,Mpc ($\mu\simeq32.3$) one might
naively think that the samples of \cite{SECM08} and
\cite{2007ApJ...661.1013L} do not suffer serious bias from missing
\underline{known} classes of SNe, as the limiting magnitude of LOSS
and other well equipped amateur searches is around $m_{\rm CCD} \sim
19$. However this is far from clear and there are arguments put
forward recently that we may even be missing events within 
10\,Mpc \citep{2008arXiv0809.0510T,SECM08}. 
The physical interpretation of the relative
frequencies and the possibility of missing events 
will be further discussed in Sections\,\ref{subsect:OTs} and
\ref{sect:overview}.

\subsection{Extragalactic stellar astrophysics from space and the ground}
\label{sec:extragstellar}
The study of individual massive stars in resolved galaxies out to 
$\sim$20\,Mpc has become fairly routine with 15\,years of 
post-refurbishment Hubble Space Telescope (HST) operations. The HST 
Key Project on the Extragalactic Distance Scale is a pioneering example 
of the feasibility of carrying out 
quantitative photometry on individual stars in other galaxies
\citep{2001ApJ...553...47F}. 
The Cepheid variables  have typical 
masses of 5-10\msol, absolute magnitudes of 
$M_{\rm V} \simeq -6$ and $(V-I)\simeq1$ \citep{1999ApJ...515....1S}.
The Key Project surveyed galaxies out to around 21\,Mpc 
identifying variable stars at $V\simeq25-26.5^{m}$ and providing
photometric precision to around 0.1-0.3$^{m}$ (in HST WFPC2 exposures
of around 2500\,s). The limit for HST images for this type of quantitative
photometry is probably around 30-40\,Mpc  \citep{1999ApJ...523..506N}.
Certainly within 20\,Mpc it is possible to resolve the brightest and
most massive stars in star forming galaxies. 
At 20\,Mpc, the 2-pixel diffraction limited 
resolution 
(at $\sim$8000\AA) of HST's Wide-Field-Channel (WFC) of 
the Advanced Camera for Surveys (ACS) of 0.1 arcsec
corresponds to 5\,pc. Thus single stars can be confused with the most
compact stellar clusters which can have diameters of between 0.5-10
pc \citep{2004A&A...416..537L,2007A&A...469..925S}. 
It is often possible to distinguish clusters from single 
stars with a combination of spectral-energy-distribution (SED), 
shape analysis and absolute luminosity \citep{2005A&A...443...79B}. 
Although the analysis methodology must be meticulous, resolving and quantifying 
the flux of individual stars at these distances is quite possible in 
HST images. If a SN is located
spatially coincident with a compact and presumably coeval stellar
cluster then it can provide a further reliable constraint on the 
progenitors age and mass. 

The largest ground-based 8-10\,m telescopes have also played a vital
role in probing the stellar content of galaxies.  Natural seeing at
the best sites on earth provides 0.6 arcsec image quality routinely in
the optical and near infra-red.  The distance limit within which
massive stars have been quantitatively studied is reduced by a factor
of approximately 6 compared to HST campaigns. The Araucaria Project
has studied Cepheids and massive blue supergiants in spirals
between 2-4.4\,Mpc
\citep{2008ApJ...681..269K,2008arXiv0808.3327G}. 
High signal-to-noise quantitative
photometric and spectroscopic data allow application of model
atmosphere and stellar wind models to determine fundamental parameters
of massive stars, even out to distances of 6-7\,Mpc
\citep[e.g. NGC3621][]{2001ApJ...548L.159B}. While the targets for
spectroscopic study are the brightest, most massive and hence rarest of 
all massive stars, these studies show that extragalactic 
stellar analysis is practicable. Stars 
may be predominately formed in clusters, but dissolution of moderate 
mass, unbound clusters on timescales of a few tens of Myrs is 
probably common place in starforming galaxies.
\citep{2006ApJ...650L.111C,2007ApJ...658L..87P}. 
 Hence the possibility 
of massive stars being resolvable in either field populations or 
resolved OB associations is relatively good. 
\cite{2006PASP..118.1626D} has studied the resolved red supergiant 
population of M81 in the NIR showing that the most massive 10-20\msol\ stars 
peak at magnitudes $M_{K} = -11.5$. Using accurate stellar photometry 
from  a 4m ground based telescope (the Canadian France Hawaiian Telescope
in this case), individual stars were easily resolved
and used to measure the recent star formation history of the disk.

\subsection{A decade of intensive searching for progenitors}
The superbly maintained and publicly accessible archive of HST 
precipitated the search for the progenitors of CCSNe discovered in 
nearby galaxies. The HST archive has become a model for other 
space and ground-based observatories world-wide. 
As described above, galaxies within about 20-30\,Mpc,
have resolved massive stellar populations in HST images and these 
galaxies are all on the SN search list of LOSS and the global 
amateur astronomy efforts.

Studies of the unresolved environments and host galaxies of SNe
started in earnest in the 1990s with \cite{1992AJ....103.1788V} and
\cite{1996AJ....111.2017V} suggesting that there was no obvious trend
for Ibc SNe to be more closely associated with giant \hii\
regions than type II SNe. 
Archive and targeted observation work with HST began after
the first servicing mission with groups looking at the resolved
stellar populations around SNe \citep{1999AJ....118.2331V}.  By the
late 1990's the HST archive, along with the highest resolution ground
based image archives, were rich enough that it was only a matter of
time before SNe exploded in galaxies with resolved massive star
populations. The cases of SN1987A and SN1993J had shown the
feasibility of progenitor classification albeit in very nearby systems
(see Section\,\ref{sect:87A93J}).  Two groups in particular began actively
searching for archive pre-explosion images for all nearby SNe. Perhaps
surprisingly the identification of progenitor stars at the positions
of these SNe was more difficult than first thought, with good images
of the II-P SNe 1999em, 1999gi and 2001du showing no progenitor
\citep{2001ApJ...556L..29S,2002ApJ...565.1089S,2003MNRAS.343..735S,
2003PASP..115..448V}. Extensive searches of the HST archive were
carried out by both groups \citep{2003PASP..115....1V,2005MNRAS.360..288M}
again with little success. Although progenitors were not discovered, the
large numbers of events and the restrictive luminosity limits were to play
an important role in investigating progenitor populations
(Section \ref{sect:IIP} and \ref{sect:Ibc}).  The first 
unambiguous discovery of a stellar progenitor in these painstaking 
searches of the HST archive which allowed the stellar 
progenitor to be quantified was for SN2003gd 
\citep{2003PASP..115.1289V,2004Sci...303..499S},  showing the expected red supergiant 
progenitor of  a type II-P SN (see Section\,\ref{subsect:03gd}) 

As these studies showed, conclusive evidence of association of a SN with a progenitor
in high resolution HST images requires differential alignment to within 
10-30 milli-arcsec, hence observation of the SN with either 
HST or adaptive optics ground-based systems is essential. There is 
a long list of misidentifications of progenitors which have used
either low resolution images or astrometry with unacceptably large
errors \citep[e.g. see][]{SECM08}. 

The discovery of the progenitor of SN~2003gd was followed by the 
hunt for progenitors for all nearby SNe in HST or ground-based images
and these are discussed in Section\,3, 4 and 5. 
\cite{SECM08} reviewed all SNe discovered within 28\,Mpc in a 10.5\,yr
period (see Section\,\ref{subsect:rates}) and found a 26\% chance that a CCSN within
this volume would have an image in the HST archive taken before explosion, 
with the SN site on the field of view of WFPC2 or ACS. 
The community have been extending this search for the precursor objects
and systems to both the Spitzer and Chandra archives
\cite[see Section\,\ref{subsect:OTs} and][]{2008MNRAS.388..487N,2008ApJ...681L...9P}

\subsection{SN impostors and their progenitors}
\label{subsect:SNimpost}
The most massive stars very likely pass through a luminous blue
variable (LBV) phase during their lifetime and the progenitors
are thought to be 
 core-H or core-He burning stars, 
ejecting their outer H (and He) envelope as they experience 
high mass-loss rates
and on the way to becoming WR stars 
\citep[see Section\,\ref{sect:mostmassive}, Figure\,\ref{fig:largeHRD} and][]{2003ARA&A..41...15M,2007ARA&A..45..177C}. During 
this phase Galactic and Local Group LBVs are known to 
show sporadic and unpredictable variability. Many show 
modulated mass-loss and variability of a few magnitudes
(commonly known as S-Doradus type variability). However 
occasionally they can undergo giant outbursts, such as the
great eruption of  $\eta$-Carina in 1843, which reached an 
amazingly bright $M_{V} \simeq -14.5$. Such energetic outbursts
have been recently discovered in nearby galaxies as optical 
transients initially identified as SN candidates. Spectroscopy
usually provides fairly unequivocal classification of these
transients as LBV eruptions and outbursts rather than SNe 
and they have been termed ``supernova impostors'' 
\citep{2000PASP..112.1532V}. 
 The identification and characterisation of 
these precursor stars will not be discussed in detail here, although
we will discuss the possibility that LBVs die in a complete 
destructive explosion in Section\,\ref{sect:mostmassive}. 
The likely LBV giant eruptions which were originally given supernova designations 
and have progenitors identified are 
are  SN~1961V  \citep{1989ApJ...342..908G,2002PASP..114..700V};
SN~1954J \citep{2005PASP..117..553V,2001PASP..113..692S} ; 
SN1978K \cite{1993ApJ...416..167R}
SN1997bs  \citep{2000PASP..112.1532V};  
SN2002kg and SN2003gm \citep{2006MNRAS.369..390M}. A complete list of 
nearby events is in  \cite{SECM08} which suggests that the rate of these
transients make up about 5\% of all SN candidates in nearby galaxies.

\newpage

\section{ Two fortuitous and surprising events:  1987A and 1993J}
\label{sect:87A93J}

Up until the establishment of voluminous space and ground-based
archives that now allow regular  searches, the hunt
for progenitor objects was confined to the closest events. 
Two SNe with clear detections of a stellar
source at the SN position are the well documented SN~1987A
and SN~1993J. Both of these events were peculiar in their own way
and surprised the SN and massive star communities by not matching the 
canonical pre-collapse stellar evolution ideas of the time.
SN1993J is most usefully discussed first as the interacting binary model
has implications for understanding SN1987A retrospectively. 

\subsection{The binary progenitor system of SN1993J}

The explosion and very early discovery of SN~1993J in M81
\citep[$d=3.6$\,Mpc,][]{2001ApJ...553...47F} provided an unprecedented
opportunity to follow the evolution of a core-collapse SN in the
northern hemisphere with modern observational techniques. The wealth
of images of this nearby spiral made a progenitor identification
almost inevitable. The photometric and spectroscopic evolution were
both peculiar, although it matched SN1987K and 
many similar examples have been found since
1993 \citep{2000AJ....120.1499M}. 
The lightcurve rose to a sharp peak only 4 days after
explosion, faded to a minimum 6 days later and rose to a secondary
peak at 25 days. The optical spectra of SN~1993J underwent
a transformation from a type II to a Ib. After 2-3 weeks the spectra
showed unusually prominent He\,{\sc i} absorption features and the
H$\alpha$ P-Cygni emission component weakened substantially 
\citep{2000AJ....120.1499M}. The lightcurve  was well matched 
with models of an explosion of a He core of mass 4-5\msol\ 
which had a residual  low mass H-envelope
(of around 0.2\msol). Three independent models of the lightcurve 
came to essentially similar conclusions for the exploding star
\citep{1993Natur.364..507N,1993Natur.364..509P,1994ApJ...429..300W}.
The low-mass, but radially extended  ($\sim$500\rsol)
H-envelope is required to 
produce the initial sharp peak in the lightcurve and 
this qualitatively accounts for the transformation
of the spectral evolution from a II to a Ib.  The
three physical models all suggested an interacting binary scenario to 
produce the  4-5\msol\ He core ; a primary star of initial mass around 15\msol\
becomes a He core-burning red supergiant which fills its Roche lobe and 
loses around 
10\msol\ during mass transfer. 

A progenitor object coincident with the position of SN1993J was rapidly 
identified and a detailed study of its $UBVR_{\rm C}I_{\rm C}$ 
spectral energy distribution
from a homogeneous set of deep images emerged. \cite{1994AJ....107..662A}
found that the SED could only be fit with two components. A red
supergiant of spectral type G8-K5I matched the $VR_{\rm C}I_{\rm C}$
colours and a blue component from either an OB association or single 
supergiant was required to account for the apparent excess in the $UB$ bands. 
The binary scenario of the progenitor being a stripped K-type supergiant
and the secondary star being an OB-supergiant was attractive 
as it could neatly 
account for the lightcurve model results, the spectral evolution and
the progenitor colours and luminosity. 
 The ground based resolution
of the best seeing 
images (1.5 arcsec at best in the blue and 1.1 arcsec in $I_{\rm C}$) 
corresponds to about 20\,pc hence the possibility of the progenitor being
embedded in an OB association was plausible. 
SN1993J remained  bright in the optical for many years due to strong nebular
lines produced by interaction of the ejecta with circumstellar material 
\citep{2000AJ....120.1499M,2007ApJ...671.1959W}
and this  
dense CSM was presumably created during the mass-transfer phase. Hence it 
required a wait of almost 10 years to search for the putative companion. 

\begin{figure}[ht]
\centerline{\psfig{file=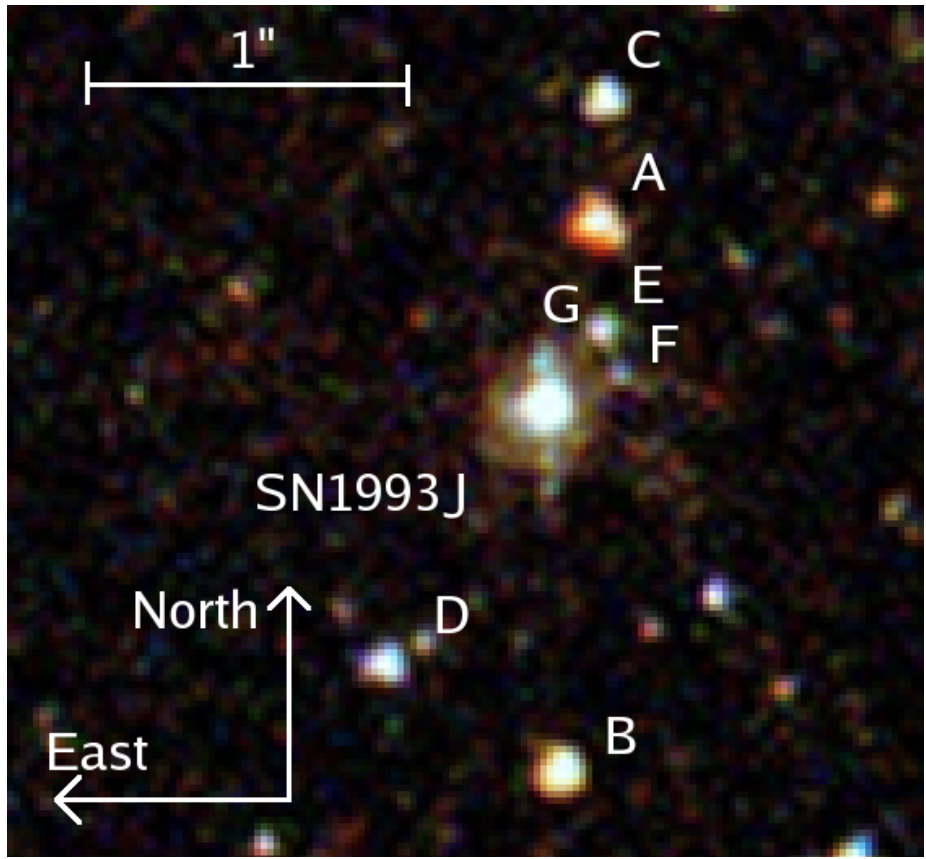,width=4.8in,angle=00}}
\caption{\small The colour combined HST ACS image of SN1993J at 10\,yrs after explosion from 
Maund et al. (2004). The progenitor of SN1993J was a bright source in the $U$ and 
$B$ bands which could either have been due to a surrounding OB-association or 
binary companion in the lower resolution ground-based pre-explosion images
of of Aldering et al. (1994). The faint blue stars E, F and G did contribute to 
the $UB$-band excess in the pre-explosion images but they cannot 
account for all the progenitors flux. A spectrum
of the SN1993J source shows H\,{\sc i} absorption lines due to a B-type supergiant
star coincident with the SN1993J remnant and this is likely the companion
to the K-type supergiant that exploded and the main source of $UB$-band flux 
in the pre-explosion images (Maund et al. 2004). 
The exposures were taken through two near-UV filters (250W, 2100
seconds and 330W, 1200 seconds) shown in purple and blue, a  blue
filter (435W, 1000 seconds) shown in green and a green filter (555W,
1120 seconds) shown in red (Image credit: 
European Space Agency and Justyn R. Maund)
}
\label{fig:93Jcolour}
\end{figure}

\cite{2002PASP..114.1322V} analysed HST  $UBVRI$ images of the site
of SN1993J taken between 1994-2001 and suggested that 4 stars lying within a 
radius of 2.5 arcsec of the progenitor position could have had enough 
flux in the $U$ and $B$ bands to account for the excess seen in the 
pre-explosion images. However this depends on how the 
fluxes are modelled and combined and  it also depends on 
how the flux of the pre-explosion source is determined. 
\cite{2002PASP..114.1322V} presented a sum of the fluxes of the
neighbouring bright stars (stars A, B, C and D in Figure\,\ref{fig:93Jcolour})
employing both a simple sum and Gaussian weighted estimate. 
As \cite{1994AJ....107..662A} used a careful 
PSF fitting method
the latter is probably  most accurate. They found that the combined 
fluxes of the neighbouring blue stars are nearly 1.4 magnitudes
fainter than the pre-explosion $B$ flux and 0.8 magnitudes fainter than the
$U$ band flux. The large uncertainties
($\pm0.5$ magnitudes), 
led \cite{2002PASP..114.1322V} to suggest that within the errors 
one could not yet claim definite evidence of further blue flux from
a binary companion at the SN position. 

\begin{figure}[ht]
\centerline{\psfig{file=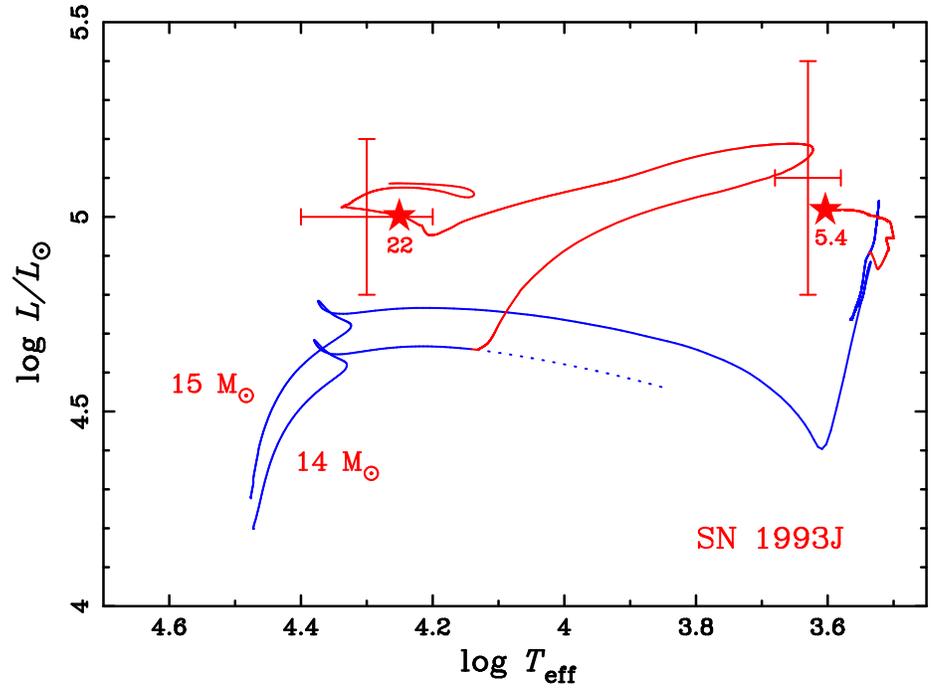,width=4.8in,angle=270}}
\caption{\small HR diagram illustrating the evolution of the binary system that 
produced SN1993J. The blue lines show the evolution of the stars before 
mass transfer, the red lines during the mass transfer phase. The numbers
give the stellar masses on the main-sequence and at the point of explosion
of the K-type primary \citep{1992ApJ...391..246P,2004Natur.427..129M}. 
}
\label{fig:93J-87A}
\end{figure}

\cite{2004Natur.427..129M} went somewhat further and imaged SN1993J ten
years after explosion with the ACS High-Resolution-Camera (HRC) on HST 
and took deep $UB$-band spectra of the SN at a moderate 
resolution (2.4\AA) with the Keck\,I telescope. The ACS image is shown in 
Figure\,\ref{fig:93Jcolour} with SN1993J  
still quite bright at this epoch ($M_{B} \simeq -8$).
They estimated the total flux contributions of the 
neighbouring sources (stars A-G in Figure\,\ref{fig:93Jcolour})
and found similar results to \cite{2002PASP..114.1322V}. 
\cite{2004Natur.427..129M} were somewhat bolder in their conclusions and stated
that the sum of the Gaussian weighted fluxes in the high resolution images
was unlikely to be able to account for the excess $UB$ light in the 
pre-explosion images. The numerical results of \cite{2002PASP..114.1322V}
and \cite{2004Natur.427..129M} are not discrepant and the conclusions drawn
differ in the interpretation of the sum of the fluxes of stars A-G. In measuring
the $B$-band pre-explosion flux, 
\cite{1994AJ....107..662A} note that their PSF fit to the $B$-band leaves
residuals to the north and south and comparing their Fig.\,1 with
the HST image in  
Figure\,\ref{fig:93Jcolour} here, it looks likely that stars A+C are the
northern residual and B+D make up the southern residual flux. Hence
 the excess $UB$-band flux 
detected at the progenitor position is not due to 
surrounding OB-stars and this now appears quite clear
 in the ACS images. 
The high signal-to-noise ratio of the Keck spectrum taken by
\cite{2004Natur.427..129M} shows distinct sharp absorption features
at the position of the H\,{\sc i} Balmer lines which were attributed
to a B-type supergiant binary companion lying coincident with the
SN1993J remnant flux. They found consistency between the pre-explosion 
magnitudes and the flux required to produce the absorption lines 
for a binary system with a B-type and K-type supergiant shown in 
Fig.\,\ref{fig:93J-87A}. 

This represents a rather satisfying picture for SN1993J in which the
unusual SN evolution is accounted for by explosion of a stripped
K-type supergiant and the detailed studies of the progenitor before
and after explosion now strongly support a binary system.  The
original mass-transfer binary model of \cite{1992ApJ...391..246P} was
adjusted, but only slightly, to better match the observations in
\cite{2004Natur.427..129M}.  Figure\,\ref{fig:93J-87A} illustrates the
pair of 15+14\msol\ stars with an initial orbital period of
5.8\,yr. The mass transfer rate is initially high (reaching a peak of 
$4 \times 10^{-2}$\msol\,yr$^{-1}$) and around 2\msol\ is lost 
to the surrounding CSM. In this model, mass transfer begins at 
the end of core He burning when the star has about ~20,000\,yrs 
to go before collapse. The extensive radio monitoring 
campaign of \cite{2007ApJ...671.1959W} suggests 
a sudden increase in the progenitors mass-loss rate
$\sim$8000\,yr before the SN and this is also supported by the 
X-ray lightcurves. This would, approximately, match the timescale
for mass lost during the mass transfer model.

Although this is a fairly consistent scenario, perhaps there
are other surprises in store, as the radio and x-ray fluxes
are now dropping indicating that the luminous interaction phase
is coming to an end. This may allow a clearer detection of the 
progenitors companion, as the \cite{2004Natur.427..129M}
ground-based spectrum and HST magnitudes were contaminated with 
the still bright remnant interaction. \cite{2006MNRAS.369L..32R}
have suggested a similar interacting binary system as the progenitor
for the IIb SN2001ig. This event  bears many similarities with SN1993J and a 
point source visible $\sim$1000 days after explosion could be 
blue supergiant (B to late F-type) companion. 

The SN that produced the Cassiopeia A remnant occurred about
1681 AD at a distance of around 3\,kpc. The detection of 
the scattered light echoes from Galactic SNe  
\citep{2008ApJ...681L..81R} now allows spectra of the 
scattered SN light (from around peak) to be collected
\citep{2008Sci...320.1195K}. This stunning look back at the
SN showed Cassiopeia A to be of type IIb, very similar to 
the time averaged optical spectrum of SN1993J. 
\cite{2008Sci...320.1195K} point out the lack of a 
detection of any viable binary companion for the Cas A 
progenitor and suggest an alternative merger scenario
\citep[e.g.][]{1995PhR...256..173N}. 
However as will be discussed in 
Section\,\ref{subsect:08ax} it is possible that some IIb come from
massive single WN-type stars.

\subsection{The nearest progenitor : SN1987A}

The most famous stellar progenitor of a supernova is
Sk$-69^{\circ}202$ which collapsed to give SN1987A in the Large
Magellanic Cloud (LMC).  \cite{1987Natur.327...36W} showed this star to be
coincident with the SN very soon after discovery and a trawl through
the photographic plate material for the LMC provided
\cite{1989A&A...219..229W} with several spectra of the star and $UBV$
magnitudes. These convincingly suggest a spectral type of B3I, a
$T_{\rm eff} \simeq 15750$ 
\citep[from the calibration of LMC B-supergiants][]{2007A&A...471..625T} and hence 
\logl$ = 5.1\pm0.1$. This star has certainly disappeared and we can now probe
deep into its core as the ejecta expands \citep{2005ApJ...629..944G,2007A&A...471..617K}.  
Extensive analysis
and discussion of the event already exists
\citep[e.g.][]{1987ApJ...319..136A,1989ARA&A..27..629A} and this
section will focus on putting SN1987A and its blue progenitor star
into context with the knowledge we now have of other progenitors. 

The detection of a neutrino burst preceding the optical explosion epoch
and the disappearance of a massive star confirms the basic theory of core-collapse. 
The main surprise in the SN1987A event was that its progenitor star was
a blue supergiant. As discussed in \cite{1989ARA&A..27..629A} and 
\cite{SECM08} the luminosity of \logl$=5.1\pm0.1$ should be compared
with the evolved He core mass, not simply the luminosity of an evolutionary
track that passes through the HRD position of Sk$-69^{\circ}202$. 
This implies a He core mass in the region $5^{+2}_{-1}$\msol, which 
can be produced from a star of initial mass in the region 14-20\msol. 
Most published tracks
of 8-25\msol\ stars still do not predict that single stars of this
mass should end their nuclear burning lives in the blue and in fact 
do not predict large numbers of He-burning (or later stage burning)
OB-type supergiants. \cite{1989ARA&A..27..629A} and \cite{1992PASP..104..717P}
show numerous examples of models which can certainly end as blue supergiants 
with appropriately chosen (and not implausible) 
parameters of mass-loss and convective overshooting. But a
consistent explanation also requires one to explain the triple ring
structure ejected by the progenitor 20,000\,yrs before explosion, 
the chemical abundances in the 
ring and also account for the properties of the supergiant population in
the LMC. 
Both binarity and rapid rotation have been proposed as explanations.

The binary model discussed for SN1993J (Figure\,\ref{fig:93J-87A}) actually 
ends with a second explosion of the blue supergiant, remarkably similar
in its predicted parameters to Sk$-69^{\circ}202$. A similar idea 
was proposed by \cite{1992A&A...260..273D} and in this case there
should be a double NS-NS system embedded in the remnant of SN1987A. 
This 
model however doesn't have a quantitative explanation for the triple ring 
morphology, although the timescales for mass ejection during 
the mass transfer phase are not inconsistent with the 20,000\,yr 
dynamical age of the rings. \cite{2007Sci...315.1103M}  invoke a 
wide binary model of a 15-16\msol\ primary and a lower mass 3-6\msol\ 
star with an orbital period of more than 10\,yrs. Unstable 
mass-transfer results in a common envelope phase and 
their 3-dimensional 
hydrodynamic model of the ejection produces a triple ring structure 
similar to that observed.

A rapidly rotating single star progenitor has alternatively been suggested as a possible 
cause of the almost axi-symmetric shape of the surrounding nebular rings. 
\cite{2008A&A...488L..37C} employ hydrodynamic calculations of the stellar wind properties 
of a 12\msol\ star which had an initial rotational velocity of 300\kms. 
 However the model star  ends its life as a 
red supergiant which doesn't match Sk$-69^{\circ}202$. 
The pre-supernova rotating model of a 20\msol\ star derived by \cite{2004A&A...425..649H}
can end its life in the blue, but 
the model star has a low hydrogen content and would probably result in a
IIb or Ib SN rather than a type II. 
There are four Galactic blue supergiants with similar circumstellar
nebulae to Sk$-69^{\circ}202$  \citep{2007AJ....134..846S}. 
An investigation into their possible binary nature, rotation rates
and photospheric abundances 
would be an important way to 
discriminate between the scenarios. 

The nitrogen abundance in the circumstellar ring found by 
\cite{1996ApJ...464..924L}
is significantly higher than the baseline LMC nitrogen 
content. The ratios of nitrogen to carbon and oxygen 
(N/C$\simeq5$ and N/O$\simeq$1 ; by number) are extremely 
high and are indicative of CNO-processed material from the 
H-burning phase having been dredged to the stellar surface and 
then ejected in the mass-loss episode that formed the ring. 
The CNO abundances in twenty-four B-type supergiants in the LMC 
were recently presented by \cite{2008ApJ...676L..29H}. The CNO ratios ranged
from $0.2 \lesssim {\rm N/C} \lesssim 8$ and 
$0.03 \lesssim {\rm N/O} \lesssim 1$. Hence the CNO 
abundances in Sk$-69^{\circ}202$ are similar to the most 
highly processed B-supergiants known in the LMC. \cite{2008ApJ...676L..29H}
showed these high abundances could be produced
by a rotationally induced mixing with a
rotation rate of $\sim$300\,\kms\ or post-red
supergiant dredge-up. At least 25\% of the highly processed
LMC B-supergiants are binaries, although their orbital 
parameters remain undetermined. 
While rapid rotation seems attractive, there 
isn't yet a single model that quantitatively explains the ring structure, 
collapse in the blue and the photospheric abundances consistently,
while also matching the properties of the OB-population of the LMC. 
The merger, interacting binary and rapid rotation models are all 
still viable and future study of the LMC B-supergiant binary population
as well as the Milky Way B-supergiants with ring nebulae seem 
promising avenues to constrain models further.

The small radius of Sk$-69^{\circ}202$ of $\sim$40\rsol\ , compared  to 
typical red supergiant radii of 500-1000\rsol\ resulted in the
distinctive bolometric and visual lightcurve of SN1987A. At the time it 
was thought that due to it being relatively faint for a type II, 
($M_{V} \simeq -15.5$ at peak) such events could have been missed within the 
$\sim 20-30$\,Mpc local volume. However it now appears that such
SN1987A-like events are indeed intrinsically rare, with 
\cite{SECM08} suggesting they are less than about 3\% of all CCSNe. 
SN1987A and SN1993J are the two most extensively 
studied SNe of modern times and neither had the expected red supergiant
progenitor expected. It appears that we have been rather fortunate, 
or unfortunate to have these explode on our door step. 
The next closest events since SN1993J 
were 2004am (M82 ; 3.3Mpc), 
2004dj (NGC2403 ; 3.3 Mpc), 2002hh and 2004et (NGC6946 ; 5.9 Mpc) and
2008bk (NGC7793 ; 3.9 Mpc). All of these were fairly normal II-P SNe
hence giving some semblance of balance to the relative 
rates of the SN types discussed in Section\,\ref{subsect:rates}. Another  
nearby event was SN1996cr which was missed at the time (in the Circinus
Galaxy ; 3.8\,Mpc) and was likely a IIn \cite{2008arXiv0804.3597B}, 
 a less common SN type. 
Additionally a number of faint, nearby transients have been discovered which have been suggested to be SNe, but their nature is currently under debate (See Section 4.5). 

\newpage

\section{The progenitors of type II-P supernovae : the most common explosion}
\label{sect:IIP}

It has been suspected for many years that the type II-P SNe are the
most common explosions, by volume, in the Universe.  The rates
compiled in Section\,\ref{subsect:rates} now quantifiably endorse this
perception.  Perhaps surprising is how rare the brighter type II-L
are. The lightcurves of II-P have generally been accepted to result
from the near instantaneous ejection of energy into an extended
hydrogen dominated envelope. Numerical hydrodynamic models
\citep{1976ApJ...207..872C} and analytic solutions of the diffusion
equation \citep{1980ApJ...237..541A} both showed that large initial
radii of order $10^{13}-10^{14}$\,cm were required. In these
calculations the energy released (in the collapse of an iron white dwarf core)
led to an expanding photosphere with velocities compatible with those
observed. For over half a century stellar evolution models have predicted
that stars between about 8-30\msol\ should begin helium core burning 
when they have expanded and cooled to become red supergiants and 
that further nuclear burning phases should occur while they 
are red supergiants.  The latter depends somewhat on the mass-loss
assigned, but standard estimates result in the end of the nuclear
burning stages being reached during the RSG stage when the stars
have radii of between 500-1500\rsol. Even the addition of
rapid rotation ($V_{\rm rot}\sim$300\kms) in the stellar models
still results in $8-22$\msol\ stars becoming red supergiants during
core He burning and beyond \citep{2004A&A...425..649H} as long
as they avoid chemically homogeneous evolution \citep{2005A&A...443..643Y}. 
The recently detected UV-flash from young II-P SNe has 
been interpreted as the shock breakout signature in a RSG progenitor
\citep{2008Sci...321..223S,2008ApJ...683L.131G}
This further strengthens the case for RSGs being the direct progenitors
of II-P SNe and may allow their density profiles to be probed in the
future. 

As the type II-P SNe dominate the rate of explosions in the nearby 
Universe it is not surprising that their progenitor population is 
observationally now the best constrained from direct detections
of progenitors or limits thereon. Images of SNe sites taken
before explosion will naturally be of variable quality in terms 
of depth, resolution and wavelength coverage. Additionally, nearby SNe
have had observing campaigns of rather variable quality and 
time coverage. Thus the total information package that is available
for a SN plus its progenitor
varies widely and the combination of high quality pre-explosion 
images with detailed observation and analysis of the SN is
the optimum dataset to physically constrain the explosion. 

The analyses of data samples of such variable quality have often
adopted subjective quality bins to describe the caliber of information
available, such as using gold and silver categories \cite[e.g. in
designating the quality of high-z SNe Ia data
sets, see][]{2007ApJ...659...98R}.  We shall group the II-P progenitor
detections into three classes to illustrate the confidence in the
progenitor detection and the quality of the data available for
characterisation of the progenitor and the SN explosion. A ``gold''
event should have enough information to estimate a colour or spectral
type of the progenitor and an initial mass.  A ``gold'' event should
also have enough monitoring data to allow the SN to be characterised.
SN2003gd, SN2005cs and SN2008bk all have unambiguous and reliable
detections ($>10\sigma$) in one or more bands. All three are almost
certainly red supergiants. Two events fall on unresolved, compact
coeval star clusters (SN2004dj and SN2004am) and  we consider these to be
gold for reasons discussed below. 
The ``silver'' events are those with a detection in one band which
is around $3-5\sigma$ or have no detailed study of the SN evolution
(SNe 1999ev, 2004A and 2004et).  The ``bronze'' are those events with no
detection of the progenitor, but with magnitude limits that set a
useful luminosity and mass constraint. The latter turn out to be very useful
as there are now a substantial number.
The results that are reviewed fall into two categories. The first are those results that are model independent, the most important of which is that the detected progenitor stars are red supergiants of moderate luminosity. However many authors have then gone one to derive quantitative luminosities and initial stellar masses. These are dependent on the stellar atmosphere models and stellar evolutionary models employed. Hence one should be careful to distinguish between results that are purely observational discoveries and those which require a theoretical model for interpretation. 

\subsection{II-P progenitors : the ``gold'' set}

\subsubsection{SN2003gd}
\label{subsect:03gd}
\begin{figure}[ht]
\centerline{\psfig{file=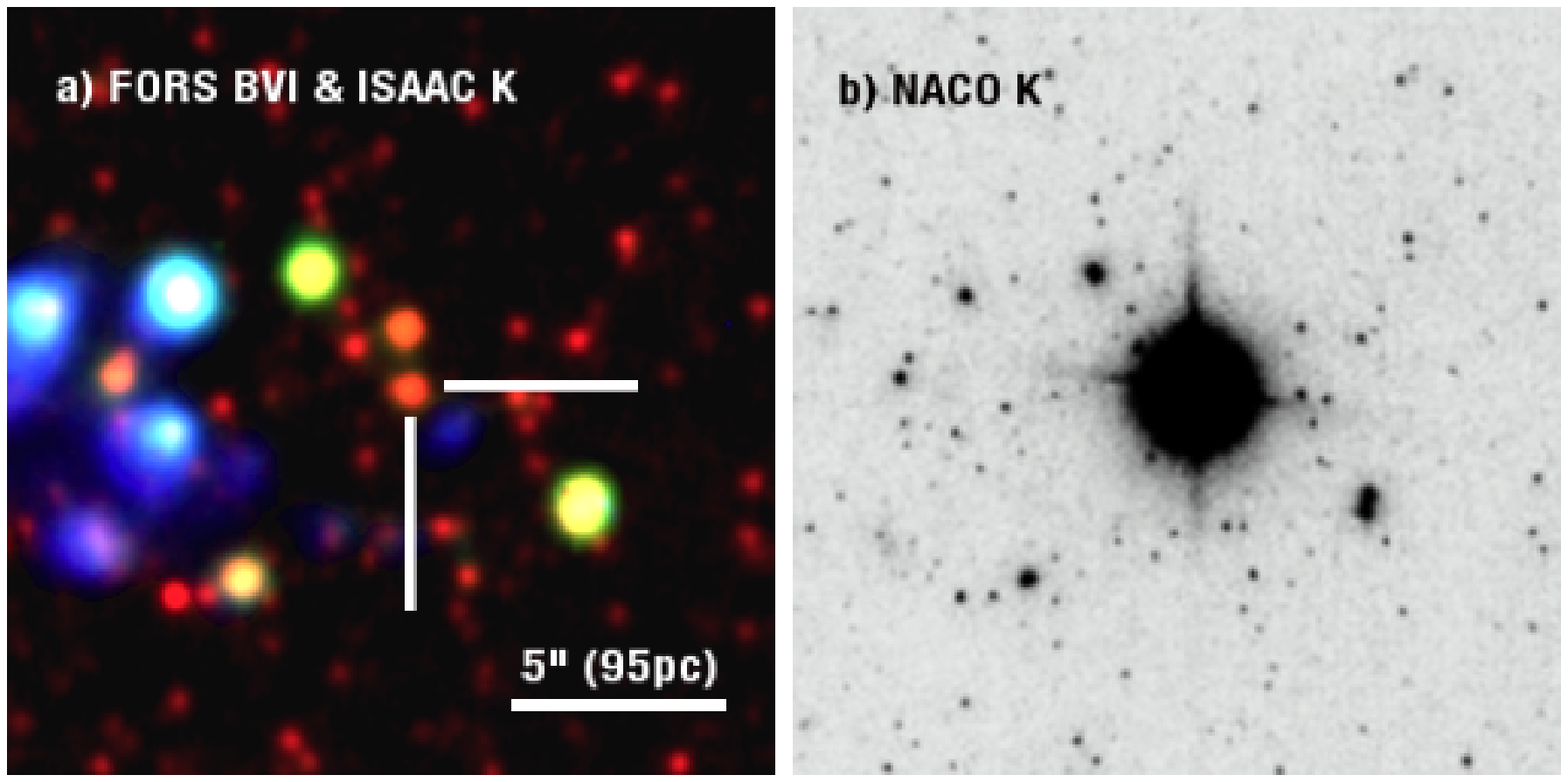,width=13cm,angle=0}}
\centerline{\psfig{file=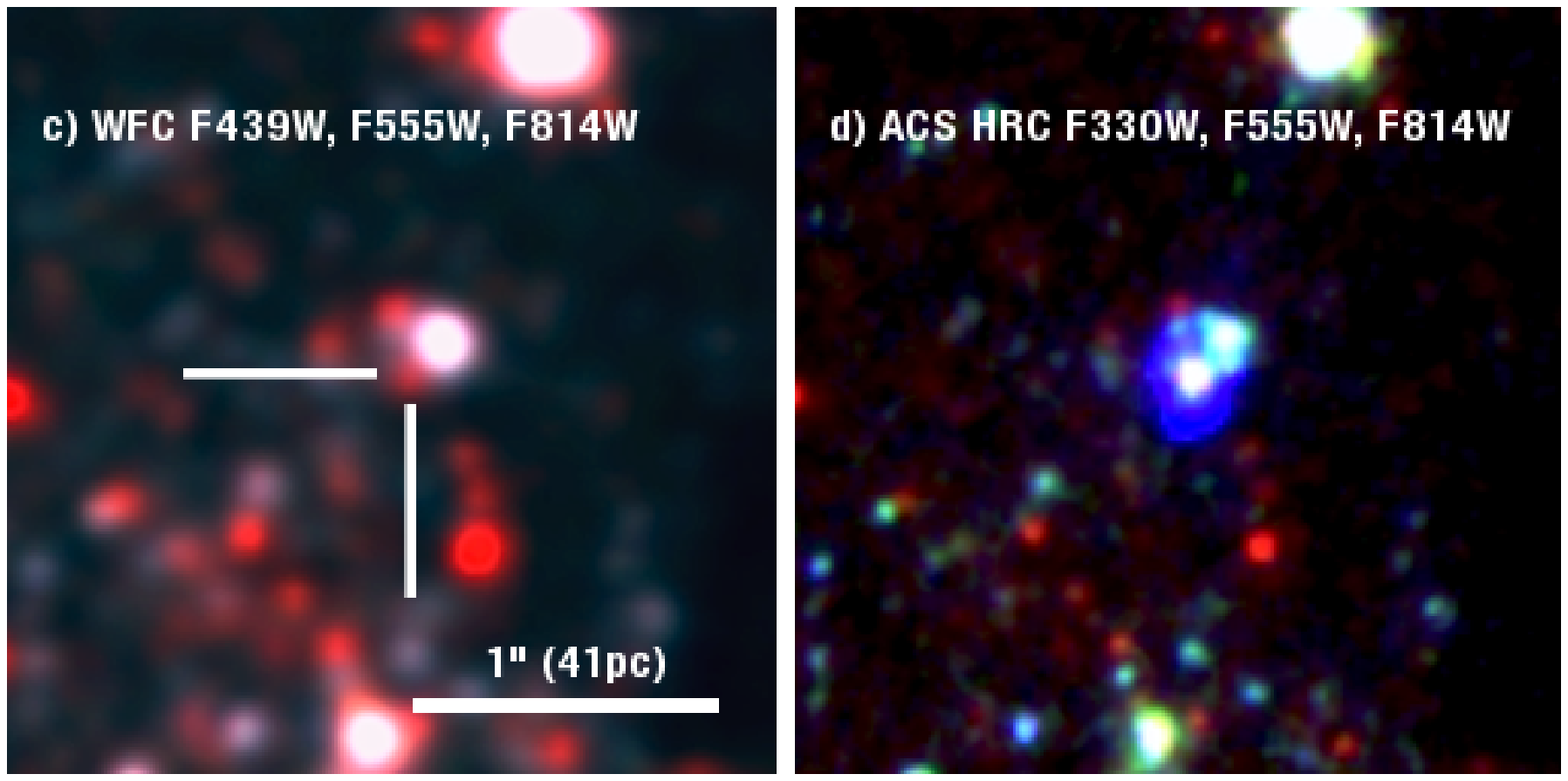,width=13cm,angle=0}}
\centerline{\psfig{file=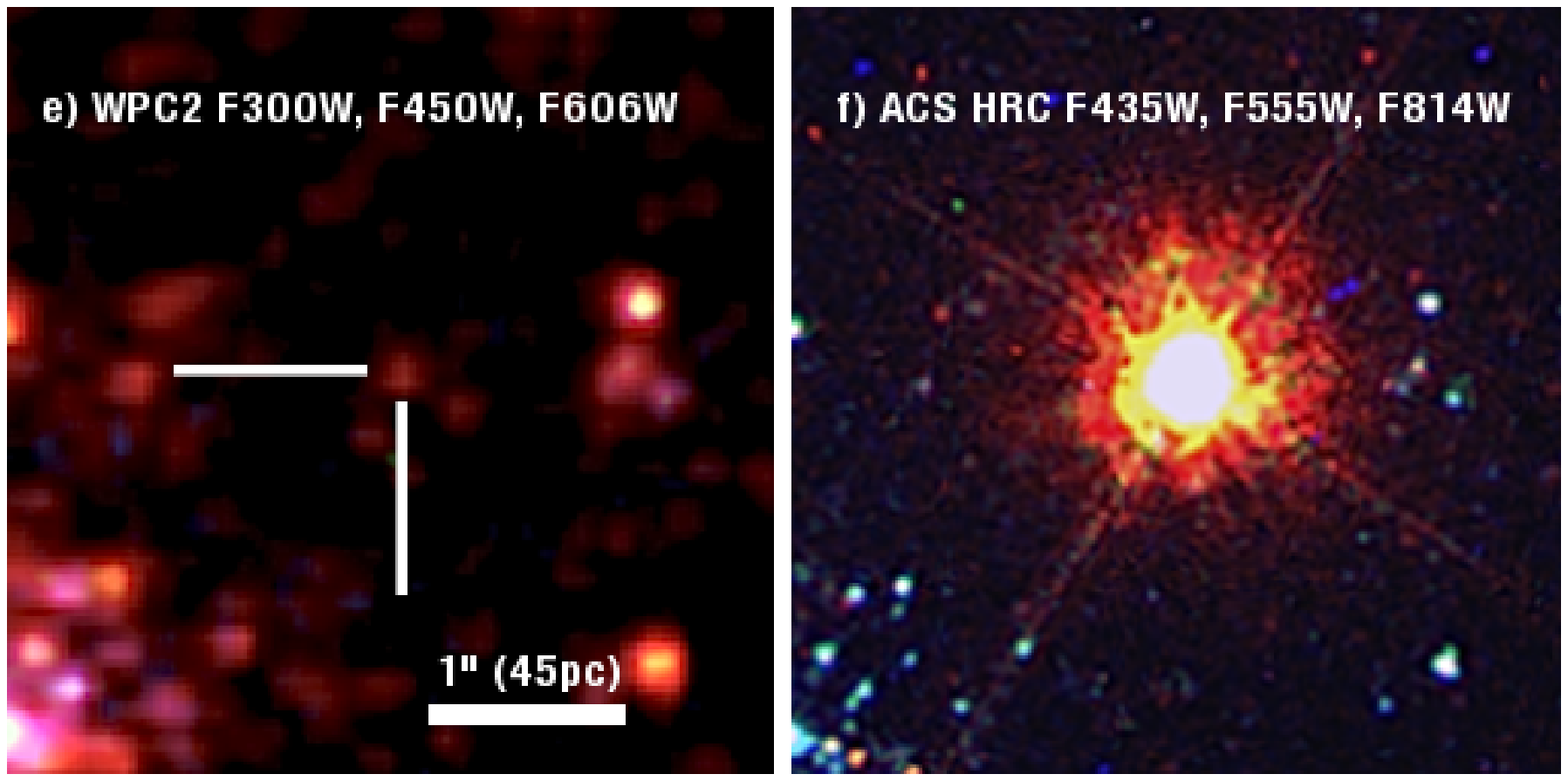,width=13cm,angle=0}}
\label{fig:IIP-images}
\end{figure}

SN2003gd exploded in the nearby face-on spiral M74 (NGC628).
\cite{2005MNRAS.359..906H} showed that it had a fairly normal plateau
luminosity and kinetic energy although it ejected a low amount of
$^{56}$Ni (around $0.02\pm0.01$\msol).  M74 had been imaged by WFPC2
on HST (3100s in F606W) and GMOS on Gemini North (480-960s in
$g'r'i'$) six to nine months before the SN explosion. A rapid attempt
to identify a progenitor using ground based astrometry isolated two
candidates  within the 0.6 arcsec error box
and the authors favoured the brighter star
\citep{2003PASP..115.1289V}.  Images of the SN with HST
showed that this single point source was coincident with the SN to
within $13\pm33$ milli-arcsec 
, which corresponds to 0.6 ± 1.5 parsecs at the distance of M74. 
\citep{2004Sci...303..499S}. The images
are shown in Fig.\,\ref{fig:IIP-images} with the progenitor identified at $V=25.8
\pm0.15$.  It is almost certain the progenitor has been identified, if
not then the progenitor must have been fainter than $V\simeq 27.1$,
which both \cite{2003PASP..115.1289V} and \cite{2004Sci...303..499S} 
note would put the progenitor mass uncomfortably below the
core-collapse limit and probably around 5\msol.  The $I$-band
magnitude of the progenitor has been estimated by both
\cite{2004Sci...303..499S} and \cite{2003PASP..115.1289V}. The value
from \cite{2004Sci...303..499S} uses
deeper, higher resolution images and
employed a deconvolution technique to estimate the flux of the
progenitor in the Gemini $i'$-band image. This resulted in $M_V =
-4.5\pm0.6$ $(V-I)_0=2.3\pm0.2$ which would imply the object is a red
supergiant within the range K5-M3Ib and the position on an HR diagram
is shown in Fig.\,\ref{fig:IIP-HRD}.
The distance to this galaxy is still, perhaps surprisingly,
not reliably determined with estimates ranging from 7.5-10.2\,Mpc
\cite[reviewed by][]{2005MNRAS.359..906H}, it would be desirable to 
establish this more reliably as the mass and luminosity estimate
of the progenitor is critically reliant on this estimate. Comparison
with the stellar evolutionary models show the progenitor is 
likely to have had an initial mass in the range 8$^{+4}_{-2}$\msol. 
The progenitor's estimated location on an HRD is similar to 
RSGs in Milky Way clusters, with the Galactic stars shown for comparison
in Figure\,\ref{fig:IIP-HRD}. The metallicity at the site of the explosion
was probably around solar. 

\subsubsection{SN2005cs}
The progenitor of SN~2005cs has been reliably identified in 
the Whirlpool galaxy M51 (NGC5194). In January 2005 the Hubble Heritage
team mapped M51 and its interacting companion galaxy 
with HST's ACS, producing a stunning  colour mosaic image
of the galaxy made from four filters (F435W, F555W, F658N, F814W). 
Rather fortuitously, SN~2005cs was discovered close to 
explosion on 2005 June 28.9. Additionally the galaxy had also been 
imaged by HST's NICMOS instrument in five near infra-red bands 
and by the Gemini-north telescope in $JHK$ with image quality of 
0.5-0.6 arcsec. Both the NIR image sets covered the pre-explosion site
of SN~2005cs providing extensive wavelength coverage for a progenitor
search. Two groups used HST to observe SN~2005cs in July 2005
to identify a progenitor \citep{2005MNRAS.364L..33M, 2006ApJ...641.1060L}. 
The two studies identified the same object in the ACS F814W images as
the likely progenitor (see Fig.\,3). Although only detected in 
one band, the limits from the other wavelengths constrain the 
progenitor to be a red supergiant, later than approximately K3-type. 
Similarly to SN~2003gd the star was quite low luminosity and low mass, 
with the two $I$-band measurements of $23.3\pm0.2$ and 
$23.5\pm0.2$  in reasonable agreement. The likely position of 
the progenitor on an HRD is shown in Fig.\,\ref{fig:IIP-HRD}, suggesting 
a mass of approximately $8\pm2$\msol\ (like SN2003gd, the nearest H\,{\sc ii}
regions in M51 display near solar metallicity). 
SN~2005cs has been followed in detail since
its explosion and is  a clear example of a low-luminosity II-P. 
(see Figure\,\ref{fig:II-P-BLCs} and Section\,\ref{subsect:56Ni}). 

The low mass of the progenitor suggests these types of explosion
come from stars at the lower mass range that can produce CCSNe. 
\cite{2007MNRAS.376L..52E} investigated the possibility that SN~2005cs was 
the explosion of a massive asymptotic giant branch star (or 
Super-AGB star) which underwent electron-capture induced 
core-collapse. They suggested this to be unlikely, from the 
restrictions on the photospheric temperature implied from the NIR
colours. 

\subsubsection{SN2008bk}
\label{subsect:08bk}
The II-P SN~2008bk exploded in the nearby Scd spiral NGC7793 at
approximately 3.9\,Mpc. This southern spiral had been extensively
imaged with ESO telescopes and deep optical and NIR images from the 
VLT provide a high quality data set for progenitor identification. 
\cite{2008arXiv0809.0206M} used the VLT NACO adaptive optics
system with the SN itself ($m_{V} \sim 13$) as a natural guide star 
to provide near diffraction limited images in the $K_{\rm S}$-band. 
Their alignment with pre-explosion $BVIJHK$ VLT images found
a progenitor star within 40 milli-arcsec of the SN position, 
 corresponding to 0.8 parsecs (Figure 4). 
The progenitor source is a strong detection 
in the $IJHK$ bands and a very red object, 
 with $I=21.2\pm0.2$ and $(I-K)=2.86\pm0.2$. 
\cite{2008arXiv0809.0206M} show the stellar SED can be fit 
by a late type M4I with $A_{V} =1$, and this corresponds to 
a red supergiant of initial mass $8.5\pm1.0$\msol. 
The metallicity of the host galaxy at the position of the explosion
appears to be low, intermediate between the SMC and LMC hence
the RSGs of the LMC and $Z=0.08$ tracks are shown in Figure\,\ref{fig:IIP-HRD}.

\subsubsection{SN2004dj and SN2004am}
\label{subsect:04dj04am}
The vast majority of CCSNe in the local Universe occur in
starforming regions of their host galaxies but perhaps somewhat
surprisingly are rarely coincident with bright star clusters
\citep{2003PASP..115....1V,2005MNRAS.360..288M}. 
Quantitatively it is probably  10\% or less.  \cite{SECM08} show 
that in their volume limited sample of twenty II-P SNe, 
only two SNe fall on compact
coeval star clusters. If these clusters are indeed coeval then a
measurement of their age gives a reasonable estimate for the
evolutionary turn-off mass and hence initial mass of the progenitor.  
SN2004dj
was coincident with the well studied compact star cluster Sandage 96
\citep{2004ApJ...615L.113M} in the nearby galaxy NGC2403.
The proximity of SN meant that it was well studied and its exploding core was found to suggest an asymmetric explosion (Leonard et al. 2006). 
 A composite
stellar population was calculated by \cite{2004ApJ...615L.113M} 
 and compared with
the cluster $UBVIJHK_{\rm S}$ observed SED. They estimated a cluster
age of approximately 14\,Myrs and hence an initial mass for the
progenitor of around 15\msol. Using different
photometry and population synthesis
models, \cite{2005ApJ...626L..89W} suggested  an age
of 20\,Myrs and a main-sequence mass of 12\msol. 
A detailed multi-wavelength study of Sandage 96 has now been carried out by 
\cite{2009ApJ...695..619V} 
after the SN faded. They determine a young age for the cluster which suggests a probable main-sequence mass for the progenitor of between 12-20\msol. 

The other example of a II-P SN originating in a star cluster is
SN2004am which is coincident with the super star cluster L in M82. 
\cite{SECM08} infer that the progenitor star had a mass
of $12^{+7}_{-3}$\msol, from the age of the star cluster of 
$18^{+17}_{-8}$\,Myrs recently estimated by \cite{2008A&A...486..165L}. 
In both these clusters there is a clear sign of a red supergiant
population either from their $JHK$ colours or the absorption 
lines in the $0.8-2.4\mu$m spectra. Although coincidences 
between SNe and compact star clusters are rare, they provide 
a valid method to estimate progenitor masses. 

\subsection{II-P progenitors : the ``silver'' set}
There are three SNe for which progenitor objects have 
been detected but the significance of the detections is either low
or more ambiguous than the gold events. 
and in one case the study of the SN itself is poor. 
The progenitor of SN1999ev
is a 4.8$\sigma$ detection in a prediscovery HST image of 
NGC4274 ($d=15.1\pm2.6$\,Mpc). It is detected at 
$m_{F555W} = 24.64\pm0.17$ or $M_{V} \simeq -6.5 \pm 0.3$
\citep{2005MNRAS.360..288M}. The sparse and mostly amateur measurements
of its photometric evolution and one spectrum suggest it
is most likely to have been a type II-P but it is not
certain. If it was a red supergiant then \cite{2005MNRAS.360..288M}
suggest a likely progenitor mass of 15-18\msol. 

There is also a probable detection (4.7$\sigma$ significance) of 
the progenitor of SN~2004A \citep{2006MNRAS.369.1303H}. 
The SN optical evolution was well studied and it 
is a fairly normal type II-P. The putative progenitor is detected
in a single filter (F814W) in an HST pre-explosion image
at $M_{I} \simeq -7.2$. The
non-detection in a fairly deep F606W suggests the progenitor was
a red star, likely a supergiant later than mid G-type
which led \cite{2006MNRAS.369.1303H} to suggest a red supergiant
progenitor of mass $9^{+3}_{-2}$\msol. 

Li et al. (2005) have claimed that the progenitor of the II-P SN2004et is a fairly massive yellow supergiant of initial mass around 15\msol . They identified the object in pre-explosion CFHT archive images of the nearby spiral NGC6946 in $BVR$ filters. This posed a challenge to well established ideas that II-P SNe came from larger radii progenitors. However, it is now clear that the object identified is not the progenitor star and is not a single yellow supergiant. Smartt et al. (2009) and Crockett (2009) show that the object is still visible at the same luminosity (in $BVR$) four years after the SN exploded. Crucially, with near-diffraction-limited Gemini NIR images, they showed that the object is a stellar cluster or association of several massive stars (see Figure 5). There is a significant difference between the pre-explosion and late post-explosion images of SN2004et in the $I-$band filter images presented by Smartt et al. (2009) which suggests that the progenitor was indeed detected, but only in the reddest optical band. The detection magnitude 
($I = 22.06\pm0.12$) and colour restriction ($R-I > 1.8 \pm 0.22$) 
led Smartt et al. (2009) to suggest it was a supergiant of spectral type M4 or later and an initial mass of $9^{+5}_{-1}$\msol.

\subsection{II-P progenitors : the ``bronze'' set}
\label{subsect:bronze}
It is routine now that the community searches 
high quality image archives for deep prediscovery images for every 
nearby CCSN discovered. But the vast majority
of SNe which have images of the pre-explosion site show no
detection of a progenitor star. 
In spite of the low rate of discovery, the sensitivity of the images can 
still set interesting restrictions on the exploding progenitor stars 
and now the large number of non-detections can be used to statistically
constrain the parent population. 

The detection of two further progenitors in Virgo cluster galaxies
was asserted by \cite{2007ApJ...661.1013L}, in which they suggested 
the identification of a red supergiant progenitor of SN~2006my and 
a yellow supergiant of SN~2006ov. However two independent studies of 
the same data have rejected these two detections. 
Using the same data \cite{leonard06my} and \cite{SECM08}
show that SN~2006my is not coincident with the \cite{2007ApJ...661.1013L} 
source. Leonard et al. (2008) estimate that the possible progenitor and SN2006my positions are not coincident with a confidence level of 96\%.  
Smartt et al. (2009) also find that the star suggested to be the progenitor of SN2006ov by  Li et al. (2007) is not coincident with the SN and cannot be confirmed as a significant detection at the correct spatial position. These two II-P events are relegated to bronze, but the upper limits derived by Li et al. (2007),  Leonard et al. (2008) and Smartt et al. (2009) are still useful.

The volume limited search of \cite{SECM08} provides a succinct 
summary of the data and information available for the progenitors 
of type II-P SNe. 
Of the 20 nearby events, 
eight are the ``gold'' and ``silver'' SNe discussed above and
twelve have no progenitor detected. Of these twelve, 
two are SN~2006my and SN~2006ov now considered
as null detections and categorized ``bronze''. 
Detection limits can be 
converted into luminosity limits by employing distance to the 
galaxy, extinction to the SN line of sight and a temperature
dependent bolometric correction 
\citep{1982ApJ...257L..63T,2001ApJ...556L..29S}. 
This defines an exclusion region in the HRD within which the 
progenitor was unlikely to lie.
This exclusion region is defined by the luminosity of a star that, if one converts its flux to a broad-band filter magnitude, would render the star detectable in the pre-explosion images.  If one assumes that the 
progenitors of II-P SNe are red supergiants 
(which seems well justified by the ``gold'' detections and 
the theory of the recombination powered plateau ; see 
Section \ref{sect:IIP})  comparison to stellar 
evolutionary models then allows an upper mass to be determined. 
Any particular mass estimate could be uncertain because of 
extinction, distance and measurement uncertainties but 
the sheer number of non-detections now appears to be significant. 

\cite{2003PASP..115....1V}  studied the HST prediscovery sites of 16 CCSNe and suggested possible progenitor candidates for a few events. However none of these have been confirmed with follow-up HST imaging. The sensitivities of the prediscovery imaging and limiting luminosities and masses tend to be meaningful for galaxies within about 20--30 Mpc (see Section 2.3); hence, the volume and time-limited sample of Smartt et al. (2009) is the most useful statistical analysis of the the masses of II-P progenitors.

\subsection{The masses of the progenitor population and the initial mass function}

The twelve upper mass limits presented in 
\cite{SECM08} (see their Table\,2) together with the eight estimates of 
progenitor masses are 
summarised in Fig.\ref{fig:IIP-imf-max}a. The mass distribution can be 
adequately fit with a Salpeter IMF of slope $\alpha = -2.35$, 
assuming a minimum mass of $8.5\pm1.0$. But this fit requires a 
fixed maximum mass of  16-17\msol. As a comparison, 
a Salpeter IMF running from 8.5 to 30\msol\ is shown and is not 
supported by the data. The lack of high mass progenitor stars
of II-P SNe is surprising. 
\cite{SECM08} have further used a maximum likelihood analysis to
estimate the best fitting minimum and maximum masses for 
the II-P progenitors. They find that the 
the minimum stellar
mass for a type II-P to form is $m_{min}=8.5^{+1}_{-1.5}$\msol\ and
the maximum mass for II-P progenitors is $m_{max}=16.5\pm1.5$\msol
(Fig.\ref{fig:IIP-imf-max}b). 
This assumes that a Salpeter IMF is appropriate for the underlying 
stellar population, although the  upper mass
limit appears robust even if the IMF slope is increased to  $\alpha = -3.00$. 
In OB associations and young clusters in the Milky Way disk and Magellanic
Clouds there is no evidence for significant deviations from 
a Salpeter type slope \citep{2003ARA&A..41...15M,2008arXiv0803.3154E}. 
The $m_{min}$ value derived appears to be a robust estimate of the minimum 
mass required to undergo core-collapse. The apparent maximum mass
that can produce a type II-P has interesting implications, which 
will be discussed further in Section\,\ref{sect:overview}.

The stellar masses and mass limits that have been derived in the 
studies discussed above are critically dependent on
theoretical stellar models. These physical models provide the 
estimate of mass from a luminosity measurement. The estimate of 
minimum and maximum masses for II-P SNe was made using the 
Cambridge STARS code \citep[see][]{2004MNRAS.353...87E}. 
The internal stellar physics in 
modern codes are fairly similar in that they employ the same nuclear
reaction rates and opacity tables. The differences are in the treatment
of mixing (convective or rotationally induced) and mass-loss. 
Both the mass-loss and rotation rates of massive stars have 
been critically linked to initial metallicity. As shown in 
\cite{SECM08} the STARS code produces model red supergiants
with luminosities very similar to the those from the rotating 
models of \cite{2004A&A...425..649H} and \cite{2000ApJ...544.1016H}. 
Thus the masses derived are likely to be similar whether rotation is 
employed or not. If mass-loss recipes beyond those adopted as
standard (or within a factor 2) are used, this could indeed affect the 
masses. Mass-loss in the red supergiant stage is particularly uncertain. 
A major uncertainty in the stellar models is the treatment of convective core overshooting. Increasing the overshooting will increase the core mass and hence its luminosity. As the surface luminosity is set by the core, the masses derived for RSG progenitors will depend on the amount of overshooting employed. This fact highlights the explicit dependence of the masses on the input physics and the stellar models. Another factor is the assumption that binaries do not play an important role in the production of II-P SNe. It is possible that the minimum initial mass could be reduced to below 8\msol\ if a lower mass star (for example around 5\msol) evolves to a higher mass through accretion.  There is no clear observational evidence for binarity in II-P SNe but theoretically the possibility remains open.

\subsection{Transients of uncertain nature : core-collapse or not ?} 

\label{subsect:OTs}
An intriguing new twist in the story of optical transients occurred in 
2007 and 2008. 
The discovery of two objects with similar luminosities, colour
temperatures and line velocities within a few months led to 
suggestions that they are physically related and that other peculiar
transients could be of the same class. 
\cite{2007Natur.447..458K} 
reported the discovery of an optical transient in M85 (M85-OT2006) and 
suggested the origin was a stellar merger, naming
the event a``luminous red nova''. 
An optical transient was discovered in NGC300 in April 2008
\citep[NGC300-OT2008][]{2008IAUC.8946....1M}
which has 
also not yet been given a supernova designation due to its uncertain nature.
Bond et al. (2009) proposed it could be outburst of a relatively massive OH/IR star rather than a true supernova explosion.
 Just 3 months earlier, a stellar eruption in NGC6946 showed similar 
photometric properties and narrow emission lines and this time was 
given a supernova designation, it is known as SN~2008S. It has been given the label of
 a supernova of type IIn based on the narrow, Balmer dominated, emission line 
spectrum. 

\cite{2008ApJ...681L...9P} and \cite{2008arXiv0809.0510T} 
have studied the pre-explosion sites of 
SN2008S and NGC300-OT2008 and found a bright mid-IR point source visible in 
Spitzer Space Telescope images (between 3.5-8.0$\mu$m) coincident with both the
eruptions. Neither progenitor was visible in deep optical images which led
the authors to suggest that these were the result of core-collapse of massive stars
which were enshrouded in an optically thick, dense dust shell. The MIR 
SED is suggestive of black body emission from the dust shell at a 
temperature of $T_{\rm dust} \sim 440-300$\,K, luminosities of 
between $\logl \sim 4.5-5.0$, and black body radii of  
$R_{\rm BB} \sim 150-520$\,AU (for SN2008S and  NGC300-OT2008 respectively). 
Stellar luminosities in this range require either evolved
massive stars (with a He core) of mass around 8-15\msol, or possibly
lower mass stars (5-8\msol) which have gone through 2nd dredge up
\citep[see Figure\,\ref{fig:IIP-HRD} and ][]{2007MNRAS.376L..52E}. 

The latter
can reach luminosities of around  $\logl \sim 4.5-5.0$\,dex and if the 
stellar flux is totally absorbed and re-emitted in the MIR they are 
plausible heating sources for the detected dust shells. \cite{2008arXiv0809.0510T}
searched multi-wavelength images 
of the Local Group spiral M33 for possible counterparts and found this
type of object extremely rare. It appears that there are fewer than 10 similar objects
 in this galaxy and they are likely extreme AGB stars. Thus a plausible 
scenario for these transients (at least SN2008S and NGC300-OT2008) is 
that they are electron-capture SNe \citep[ECSNe;][]{1987ApJ...322..206N,2006A&A...450..345K}. 

 The progenitors would be 
super-AGB stars, having undergone 2nd dredge up and carbon ignition, 
and collapse of their O-Mg-Ne cores is triggered by electron capture
before Ne ignites \citep{1984ApJ...277..791N,2008ApJ...675..614P}. 
Various groups are monitoring SN2008S and NGC300-OT2008
transients intensely and conclusions as to the explosive nature of
the two transients will be forthcoming soon. 
Three ways to provide evidence for 
the ECSNe scenario are the detection of a $^{56}$Ni 
decay phase, possible broad-lines from intermediate mass element ejecta in the 
very late time spectra  and the disappearance of the progenitors in future observations. 

There is no Spitzer source at the position of M85-OT2006 in an image
from 2004 but \cite{2008arXiv0809.0510T} 
note that the post-explosion MIR evolution may be comparable
to SN2008S and  NGC300-OT2008, hence suggesting a common origin. 
Whether or not all three transients are really of the same nature and 
whether or not they are ECSN from dust obscured super-AGB stars still remains
to be confirmed. The alternative scenario put forward by \cite{2007Natur.447..458K}
is that M85-OT2006 is the result of a violent merger of a low or intermediate 
mass star with a more massive primary or a compact remnant. This is 
still a viable possibility for M85-OT2006 and also for the 
other two. 
A full comparison
of the energetics and kinematics of all three events (and also possibly SN~1999bw ; see
Thompson et al. 2008) will guide future discussion.

\newpage
\section{The progenitors of Ibc supernovae}
\label{sect:Ibc}

The simple fact that Ibc SNe do not, on the whole, show evidence for 
hydrogen ejected at  velocities similar to the intermediate mass
elements is convincing evidence that the exploding star did not
have a hydrogen atmosphere. It is likely that some Ib SNe do show
evidence of hydrogen absorption features in their early photospheric
spectra \citep{2002ApJ...566.1005B}
and there is almost certainly 
a continuum of hydrogen
line strengths between the classic Ib SNe (with no sign of H)
and the IIb \citep{2006A&A...450..305E}. 
The progenitors of Ib and Ic SNe have been proposed
to be massive Wolf-Rayet stars  \citep{1986ApJ...306L..77G}
as these are massive evolved stars that have shed most, if not
all, of their hydrogen envelope. 
An alternative scenario is that 
the Ibc SNe progenitors are stars of much lower initial 
mass in close binaries which have had their envelopes stripped 
through interaction \citep[Roche lobe overflow, or common envelope 
evolution;][]{1992ApJ...391..246P,1995PhR...256..173N}. 
This section will review the 
evidence from direct searches for progenitors of Ibc SNe within
about 30\,Mpc and we will include the IIb SNe in this discussion
as they have also been stripped of  much of their hydrogen atmosphere.

\subsection{Searches for Ibc progenitors}
\label{subsect:Ibcdirect}

There are 10 SNe classified as Ibc which have deep pre-explosion images
available and none of them have a progenitor detected. \cite{2005MNRAS.360..288M}
and \cite{2005ApJ...630L..33M} attempted to use a combination of evolutionary
models of single WR stars and model spectra to constrain the physical 
parameters of the progenitors. \cite{2007MNRAS.381..835C} also discussed this approach for 
SN~2002ap but the uncertain and variable bolometric correction 
of WR stars makes it 
difficult to determine restrictions on mass. WR stars in the LMC and Milky Way
show highly variable broad-band magnitudes with little direct correlation
with current (or initial) mass. \cite{2005ApJ...630L..29G} have preferred a 
simpler comparison of their magnitude limit for the progenitor of SN~2004gt
with known WR populations. \cite{2003PASP..115....1V} carried out a similar 
comparison for several Ibc SNe. 
Figure\,\ref{fig:WRIbc} shows the broad-band 
magnitudes of WR stars in the LMC with a comparison of the limits for all
the Ibc progenitors with HST pre-explosion images (or deep CFHT images in the 
case of SN2002ap). The deepest limit is for the Ic SN2002ap in which there
is no detection of a progenitor star to a limit of 
$M_{B} \geq -4.2\pm0.5$ 
and
$M_{R} \geq -5.1\pm0.5$. For this event and 
any other individual SN in Fig\,\ref{fig:WRIbc} the 
magnitude limits cannot rule out a massive WR star progenitor. However
lets make a hypothesis that the progenitor population of all Ibc SNe are
massive WR stars as we see in the Local Group (and that the LMC
luminosity distribution is a fair reflection). Then we can ask, 
what is the probability
that we have not detected any of the 10 progenitors simply by chance. 
A simple probability calculation would suggest the probability is
11\% if one assume that the likely Ib progenitors are WN stars and 
Ic progenitors are WC/WO stars. 
Thus we conclude, at 90\% confidence level that the hypothesis is false and 
the massive WR population we see in the Local Group cannot be
the only progenitor channel for Ibc SNe. 
The implication is then that some of the population come 
from lower mass stars within interacting binaries and 
how this compares with the rate of Ibc SNe will be discussed below. 
The following two sections discuss interesting events in which a 
possible WN progenitor has been detected and a possible host cluster
has been identified. They represent two of the best
opportunities for characterising the local IIb-Ib-Ic populations.

\subsection{SN2008ax : a WNL progenitor of a IIb or a binary in a cluster ?}
\label{subsect:08ax}
A detection of a point source coincident with a IIb SN has been 
reported for SN2008ax in NGC4990. This event had a bolometric lightcurve 
almost identical to SN1993J apart from no detected shock breakout and
the early explosion phase was well enough observed for this to be a 
robust conclusion \citep{2008MNRAS.389..955P}. 
The strong H$\alpha$ absorption
feature in the spectrum faded rapidly and by 56 days nearly all traces
of hydrogen had disappeared from the spectrum which became He dominated. 
\cite{2008arXiv0805.1913C} showed that the SN was coincident to within 
22 milli-arcsecs of a bright point-like source detected in three HST
bands (F435W, F606W and F814W) in pre-explosion WFPC2 images. 
Using a distance of 9.8\,Mpc and extinction of $E(B-V)= 0.3$, 
\cite{2008arXiv0805.1913C} estimated absolute magnitudes of 
$M_B = -7.4 \pm 0.3$, 
$M_V = -7.3 \pm 0.3$,
$M_I = -7.8 \pm 0.3$. 
A single supergiant SED cannot be fit to these colours and 
\cite{2008arXiv0805.1913C} show that it is difficult to come up 
with a binary system which has a combined colour matching 
the observed
 and  consistent luminosities to explain the evolutionary 
path to explosion for the more evolved star. The progenitor could have been 
a binary, similar to that proposed for SN1993J, but with 
additional flux within the PSF from other neighbouring stars. 
The object is consistent with a single PSF, but at a distance of nearly
10Mpc, the PSF width corresponds to about 6pc. 
\cite{2008arXiv0805.1913C} propose that the magnitudes are 
similar to WN and WNL stars in the LMC and M31. The progenitor
of SN2008ax would be one of the brightest of this population but
its colours are quite consistent with it being such a stripped massive
star and possibly of initial mass between 25-30\msol. 
Hence this remains
the only possible direct detection of a WR star as a SN progenitor
and the comparison models shown in \cite{2008arXiv0805.1913C} show
reasonable agreement with the final position of the progenitor in 
colour magnitude diagrams. When the SN fades we shall see if this
object disappears, which it should if a massive WR star origin is 
correct, or if the ``binary within a cluster'' scenario is true.
The SN was not a Ibc, but a IIb in which clear evidence of hydrogen
was seen although the transformation to a Ib was more rapid than
that seen in SN1993J. 
The lack of a strong shock breakout 
is suggestive that the stellar radius was much smaller
than the extended (but H-deficient) K-type supergiant proposed
for SN1993J, hence suggesting a compact WN star could be viable. 
The \cite{1993Natur.364..507N}
model of SN1993J required an extended, but low mass H-shell to 
reproduce the shock breakout and naked He-cores produced the 
secondary rise well without the initial luminosity peak.

\subsection{SN2007gr : possible mass estimate from host cluster properties}
As discussed above in Section\,\ref{subsect:04dj04am}, if a SN is 
spatially coincident with a coeval compact star cluster one can 
probably assume membership. Hence a measurement of the cluster 
age and turn-off mass for a coincident Ibc SN is potentially very 
interesting. \cite{2008ApJ...672L..99C} show that the Ic SN2007gr
lies on the edge of a bright source, 6.9pc from its nominal centre and 
that the bright source is probably a compact cluster. The pre-explosion
HST images are not of wide enough wavelength coverage to determine
a unique age for the cluster, or indeed confirm for certain that it is not
an extremely bright single supergiant. 
A future combined optical and NIR SED of the possible host cluster could
give a robust age. \cite{2008ApJ...672L..99C} suggest that this could  
distinguish between two likely turn-off ages of around 7 and 25\,Myr. 
In principle it may be possible to favour a massive single WR star 
 (around 30\msol) or an interacting, lower
mass binary (around 10\msol) from the cluster age.

\subsection{The rate of Ibc SNe and interacting binary stars}
\label{subsect:rateIbc}

The relative frequency of discovery of SNe Ibc is strongly suggestive that 
at least a fraction come from interacting binaries. The $N_{\rm Ibc}/N_{\rm II}$
ratio (discussed in Section\,\ref{subsect:rates}) is $0.4\pm0.1$ at metallicities
of around solar. If we were to assume that this is simply due to higher mass stars
producing Ibc by becoming WR stars then the formation of a WR star must occur at
initial masses of about 16\,\msol\ and above. This is much too low
to be consistent with initial masses for WR stars in the Local Group. 
In the Galaxy and LMC clusters, the 
turn off mass to produce WN stars is at least 25\msol\ 
and probably closer to 35-40\msol\ to produce WC stars
\citep{2003ARA&A..41...15M,2007ARA&A..45..177C}. Also the observed 
mass-loss rate of 16-20\msol\ stars would be somewhat too low to produce
WR stars in evolutionary models which adopt these \.{M} values
\citep[see][]{2003ApJ...591..288H,2004A&A...425..649H,2004MNRAS.353...87E,2007ARA&A..45..177C}.

The high rate of 
Ibc SNe was recognised as a problem in the 1990's 
and interacting binaries were suggested as a common channel
\citep{1995PhR...256..173N}. 
\cite{1992ApJ...391..246P} calculated that 15-30\% of all massive stars (with
initial masses above 8\msol) could conceivably lose mass to an interacting
companion and end up as a helium star. They assumed a fraction of stars in 
binary systems which are close enough to interact of about a third. 
This latter fraction is still uncertain and recent results suggest it
could be more than 60\% \citep{2007ApJ...670..747K}. The lack of 
detection of any massive WR progenitors would point towards the binary 
channel being a common cause of stripped, evolved stars at their life's
end. All that is required is that the primary star in the system is
more massive than about 8-10\msol, a companion of a few \msol\
and  an orbital period  less than around 100 yrs. Such systems
are not uncommon in our galaxy,  for example 
V Sagittae, WR 7a and  HD45166 are all binary systems with a H-deficient
primary that has probably lost its mass either through 
Roche-lobe overflow or common envelope evolution. But whether or
not they will explode as type Ibc SNe and how
common they are by volume are both unanswered questions. 
If they are common progenitors of type Ibc then they should
nearly be as common (within $\simeq$30\%)  as 
evolved 
massive stars (blue and red supergiants). Perhaps the final mass-transfer that 
strips the core occurs very close to the end of nuclear
burning (in the last $\sim10^{4}$\,yrs) 
and thus the phase lasts such a short time that they are
rare objects. Alternatively \cite{1995PhR...256..173N} has proposed
that common envelope evolution in binaries can result in progressively
severe stripping of the envelope of the primary, leading to a 
sequence of II-L, IIn, IIb, Ib, and Ic.

There are theoretical arguments that massive WR stars
collapse to form black holes and that, at solar metallicity and 
below, they do not form bright SN explosions. 
In related papers 
\cite{2003ApJ...591..288H} and \cite{1999ApJ...522..413F}
put forward the idea that at around solar metallicity a 
star which is massive enough to shed its envelope 
through radiatively driven winds ($\sim30-60$\msol\ with their
adopted mass-loss recipe) ends up with a core
mass that is too large to form a neutron star. When a 
black hole is formed, fall back means little $^{56}$Ni
is ejected and an electromagnetically weak explosion follows. 
By extrapolating mass-loss rates above solar 
metallicity they suggest that the mass-loss rate 
could be high enough so that stars 
with ZAMS mass $M_{\rm initial}>25$\msol\  
produce the canonical core-collapse to a neutron star and 
successful neutrino driven shock. This is course still 
uncertain as mass-loss at high metallicities remains
unconstrained as do stellar abundances.  
\cite{2007PASP..119.1211F} put forward the idea that
{\em all} bright Ibc could conceivably come from
interacting binaries, and massive WR stars 
could be collapsing quietly to black hole holes
with no visible explosion. 
\cite{2008MNRAS.384.1109E}
illustrate that by mixing single stars and interacting 
binaries in massive stellar populations
they can reproduce the Ibc ratio at solar metallicity
and get a lower value of $N_{\rm Ibc}/N_{\rm II} \sim 0.1$
at 0.3\zsol, as suggested in the surveys discussed in 
Section\,\ref{subsect:rates}. 
This is further encouragement for the observers to improve
the metallicity determinations of nearby SNe environments.

\subsection{The environments of type Ibc SNe}

A strong argument that Ibc SNe actually do come from stars of higher masses 
than type II-P is their association with \hii\ emission and
areas of high stellar surface brightness in their host galaxies. 
An early study of the proximity of the Ibc and II SNe with 
\hii\ regions suggested the degree of association was not 
markedly different  \citep{1996AJ....111.2017V}. However a factor of 
two increase in the numbers of SNe available suggest differences
are now discernible.

\cite{2008arXiv0809.0236A} show that the positions of SNe Ic 
in late-type galaxies tend to trace the H$\alpha$+[N\,{\sc ii}]
line emission. This contrasts markedly with the locality of SNe II, 
which are not, on the whole, associated with \hii\ regions. The
SNe Ib also show a higher degree of association with the 
H$\alpha$+[N\,{\sc ii}] emission than the SNe II, although
somewhat less than for the Ic. As \hii\ emission requires a 
young population of ionizing sources (O-stars) the implication is 
that the SNe Ic come from a younger population of progenitors
than SNe II (with the Ib in between). 
\cite{2007arXiv0712.0430K} reach a similar conclusion in 
finding that the SN Ic tend to fall on areas of higher
surface brightness than the SNe Ib and II, from 
surface brightness maps in SDSS host galaxies.
The statistics from these studies are impressive, with 
69 (type II),  11 (Ib), 24 (Ic)  from  \cite{2007arXiv0712.0430K} and 
100, 22, 34  from  \cite{2008arXiv0809.0236A}. The case for
an increasing mass range for progenitors of SNe II-Ib-Ic 
is supported by both these studies. 
However as bright \hii\ emission
and integrated continuum light is indicative of 
high stellar surface density and high specific starformation rates, 
it is also likely to trace cluster and OB-association localities. 
\cite{2008A&A...477..147C} point out that the binary fraction
in field stars is lower than that found in stellar clusters
and OB-associations. While this is still not definitively 
proven, perhaps there is a propensity for a higher
binary fraction in these regions. One might then imagine
that these regions could conceivably produce higher numbers of 
Ibc SNe.

\subsection{Ejecta masses from SNe Ibc and GRB related SNe}
With the lack of detection of a progenitor of a Ibc event, the only
other way to determine a stellar mass is from modelling of the lightcurve
and spectral evolution. The type Ic SNe have been subject to intense
scrutiny recently due to their link with long-duration GRBs (LGRBs)
with ejecta masses now determined for nine Ibc SNe
\citep[][and references therein]{2006Natur.442.1018M,2008MNRAS.383.1485V}. 
The lowest of these are 1994I, 2002ap and 2007gr with ejecta masses
between 1-2.5\msol. The mass of the remnant left is then critical 
for an estimate of the CO core that exploded. 
If we assume a canonical mass of 1.5\msol\ for a neutron star 
remnant, then the CO core masses of these objects would be 2.5-4\msol.
These are lower than typically found for the current 
masses of WC stars in the Galaxy and LMC \citep{2002A&A...392..653C}
of between 7-20\msol.  With total energies of 
around $1-4 \times 10^{51}$\ergsec, these are the least energetic of 
the Ic SNe that have been modelled. The likely scenario is then that
they were not single, massive WC stars but that the
CO core of this low mass was formed in an interacting 
binary. In these models a CO core of 3-5\msol\ corresponds to 
a primary of initial mass 
around 8-15\msol. Although only a few of the nine have low masses, this
is due to the high energy events being preferentially selected for 
detailed modelling and is not a reflection on the relative 
rates.

The more energetic events, in terms of their kinetic energy and 
bolometric lightcurves, indicate higher model
ejecta masses. The LGRB related SNe (SN1998bw, SN2003dh, SN2003lw)
have estimated ejecta masses of 8-13\msol, while the energetic
SN2004aw and SN2003jd (which lack detected LGRBs) were
calculated at 3-5\msol
\citep{2006MNRAS.371.1459T,2006ApJ...645.1323M,2008MNRAS.383.1485V}.
Adding a minimum of 1.5-2.5\msol\ for a NS/BH remnant 
would suggest reasonable agreement between the progenitor CO
core mass and LMC WC stars. 
Although systematics may affect the masses determined by
the lightcurve modelling technique and they are not yet 
observationally confirmed with an independent method, 
it does appear that the 
relative difference in the shapes of Ic SNe are due to 
an increasing ejecta mass and an increasing mass of the CO 
star which exploded. 
The most energetic of these are associated with GRBs. 

\cite{2004ApJ...607L..17P} suggested that the rate of 
energetic broad-lined Ic SNe is similar to the rate of LGRB
which might indicate that most (or all) energetic Ic SNe 
produce GRBs. This assumed that $\sim$5\% of all Ibc SNe
were energetic Ic and this is supported in the volume limited
numbers of \cite{SECM08}; of 27 Ibc only one (2002ap) would
qualify as a broad-lined Ic. As \cite{2004ApJ...607L..17P}
point out, that the observed rate of production WR stars in galaxies
(from stars with  initial masses $>40$\msol) far outweighs 
(by a factor of $\sim10^{2}$) the broad-lined Ic SN rate.
Thus it is certain that not all WR stars produce broad lined Ic SNe. 
If we have reason to believe that the normal
Ibc population do not, on the whole, come from 
massive WR stars (see Section\,\ref{subsect:Ibcdirect}) then 
what is the fate of these stars ? A further complication
is that the observed WC/WN ratio is between 0.1 (at SMC metallicity)
and 1.2 (solar metallicity ; see Crowther 2007 and Massey 2003) but 
the Ic/Ib rate is $2\pm0.8$ (Section\,\ref{subsect:rates}). Either the
WN phase is a transient evolutionary phase for WR stars, or binary systems
significantly alter the Ic/Ib ratio significantly.

In summary the observational evidence supports the ideas that a significant fraction
of Ibc SNe coming from interacting binaries in which the primary that
explodes has a mass lower than what is usually associated with 
evolution to the massive WR phase. This is supported by the lack
of progenitor detections and the low ejecta masses for the least
energetic Ic SNe. Although some objects with low ejecta masses
clearly have high kinetic energies (SN2002ap for example).  
However the birth places of Ibc SNe suggest that 
the Ic SNe, when taken as a population, come from noticeably 
younger (or denser) regions than the type II SNe. This could imply that they
have appreciably higher initial mass. The ejecta masses of the 
most energetic events would also indicate they could be from
 massive single stars that form WRs. Hence there are likely 
two channels at work. The relative contribution of each remains
to be determined and the exact relation between core-mass, 
\nick\ production, kinetic energy and compact remnant is an
area for future study.

\newpage
\section{ The fate of very massive stars}
\label{sect:mostmassive}

The most massive stars known in the Local Group are LBVs which are
evolved blue stars with strong winds and luminosities between 
$5.5<$\logl$<6.0$ 
\citep{1994PASP..106.1025H}. 
The most extreme have evolutionary masses in the range
80-120\msol. Their position on the theoretical HRD and comparison with
evolutionary tracks implies that they are either core H-burning or
He-burning stars which have evolved from the main-sequence
(Figure\,\ref{fig:largeHRD}). 
Evolutionary scenarios based on stellar evolution theory and
observational inferences from massive stellar populations in the Local
Group have generally implied, at least up until now, 
that they are likely to lose their H and He envelopes and
end up as WR-stars \citep{1994A&A...287..803M,2003ApJ...591..288H,2003ARA&A..41...15M}. 
Recently \cite{2007A&A...475L..19L} 
have proposed that some very massive stars may retain at least part of their H-envelope until their deaths. 
Although radiatively driven mass-loss occurs during
the LBV phase and in the massive O-star progenitor phase, the current
measurements of rates are  too low to completely drive off the H and
He atmospheres, particularly when wind clumping effects are
considered \citep{2006ApJ...645L..45S}. 
They can lose several 
solar masses of material in short and sporadic eruptions \citep{1994PASP..106.1025H}
and the physical cause is not well understood
\citep{1990A&A...237..409P, 2006ApJ...645L..45S,2004ApJ...615..475S}. 
Very large ejecta masses of around 10\msol\ in these sporadic outbursts have been suggested along with the idea that only super-Eddington continuum winds or hydrodynamic explosions could be the cause \citep{2006ApJ...645L..45S}. 
Thus the ultimate fate of these most massive stars has been 
uncertain. Their core masses at the end of evolution would
suggest that they are likely to form black holes, if the 
core collapses in a similar way to lower mass objects
\citep{1999ApJ...522..413F,2002RvMP...74.1015W,2003ApJ...591..288H,2006NuPhA.777..424N}. 
Several 
unexpected and extraordinary discoveries in the last three years
have opened up the debate on the physical process that governs
the death of these stars. The core-collapse mechanism 
struggles to explain their nature and novel explosion physics
has already been developed. 

\subsection{SN2005gl : a very massive star}
Although Sections\,\ref{sect:IIP} and \ref{sect:Ibc} have concentrated
on searches for progenitors in galaxies closer than about 30\,Mpc, 
studies of the environments of a small number of SNe at larger
distances (40-100\,Mpc) were being carried out \citep{2003PASP..115....1V}. 
The possibility of even HST images being sensitive to 
individual stars relied on locating
very bright and hence very massive progenitors. A remarkable 
discovery
by \cite{gal-yam_leonard} shows that a star which is likely one of
the most massive and luminous stars we know exist exploded to 
produce a IIn SN.  When SN2005gl was discovered, 
\cite{2007ApJ...656..372G} located an HST image of the host galaxy
NGC266 taken in 1997. Images in two filters were available (F547M: medium
width $V$-band and F218W: UV band) and alignment with a
high resolution image taken with the Keck laser guide star AO system
showed a bright point source (only in the F547M band) 
coincident with the SN.
\cite{gal-yam_leonard} then showed that the star has disappeared
in subsequent HST images with the same filter 
(see Figure\,\ref{fig:05gl}).  The progenitor 
was observed with $M_{V} = -10.3$ and assuming a zero bolometric
correction this implies a luminosity of \logl=$10^{6}$. The 
only stars known locally of this luminosity and visual magnitude
are the luminous, classical LBVs such as AG Car, AF And, P Cyg 
and S Dor \cite[see][for a summary of LBV luminosities, and 
Figure\,\ref{fig:largeHRD}]{2004ApJ...615..475S}. 
SN2005gl was a relatively bright SN IIn which shows distinct
evidence of the SN ejecta interacting with a circumstellar
shell (Figure\,\ref{fig:05gl}). 
The narrow H$\alpha$ line in the spectrum 8 days after discovery
suggests the existence of a shell of H-rich gas with 
an outflow velocity of around 450\kms. The later spectra
at days 58 and 87 show the broader profile of the SN ejecta
moving at around 10,000\kms. From these spectra and the lightcurve, 
\cite{gal-yam_leonard} estimate that the progenitor 
lost a modest amount of mass ($\sim$0.03\msol) to create the
circumstellar shell but that the lack of an extended plateau probably
points to it having shed a considerable amount of its H-envelope
before explosion.

\subsection{SN2006jc : a giant outburst followed by core-collapse}

The first discovery of a bright optical transient spatially
coincident with a subsequent luminous supernova was reported by
\cite{2007Natur.447..829P}.  The SN2006jc was preceded, two years
earlier, by a sharply decaying outburst that reached $M_{R} \simeq -14.1$ and
was detected for only a few days.  The outburst magnitude and fast
decline is similar to the giant outbursts of some LBVs. These
outbursts have been recorded in the Galaxy ($\eta$ Car and P-Cygni)
and in the nearby Universe (Section\,\ref{subsect:SNimpost}), but they
have generally been thought to be associated with a mass ejection
event in which somewhere between a few tenths and few solar masses are
ejected. As the known LBVs, which have exhibited this behaviour, still
retain their H-envelopes, the material is normally H and He
rich.  \citep{2007ApJ...657L.105F} and \cite{2007Natur.447..829P} showed that
the high velocity ejecta spectrum of SN2006jc is more like 
a type Ic, with intermediate mass elements O, Mg, Ca (and possibly Na and Si)
exhibiting outflow velocities of 4000-9000\kms. Strong  He lines
are persistent, but with a lower velocity of around 2000\kms\ and
weak H is detected at later times. 
The narrow He\,{\sc i} lines
are circumstellar and this material was ejected from the star 
in the recent past, although not necessarily in the 2004 outburst.
This led to the conclusion that the exploding
star was a WC or WO star embedded within a He rich circumstellar 
envelope \citep{2007ApJ...657L.105F,2007Natur.447..829P,2007arXiv0711.4782T}. 
 The outburst in 2004 had a peak luminosity of 
at least \logl$\sim7.5$ and total integrated energy  over 9 days of
$\gtrsim10^{47}$ergs. This is similar to the known 
outbursts of high luminosity LBVs \citep{1999PASP..111.1124H}, but all of these
still retain significant hydrogen and helium atmospheres. 
LBV stars are often helium enriched but are not completely
deficient in hydrogen. The progenitor of SN2006jc was a
CO core explosion which raises unanswered questions about the
outburst. \cite{2007arXiv0711.4782T} calculate a mass for the 
WC/WO star of 6.9\msol\ and an initial mass of around
40\msol\ on the main sequence. 
Such energetic outbursts  have never been associated 
with WR stars and this may the first observed example of 
a star transitioning from the LBV phase to the WR phase
through sporadic mass ejections. It may be that the $10^{47}$ ergs
outburst ejected the last remnants of its outer He layer 
\citep{2007arXiv0711.4782T,2007ApJ...657L.105F,2007Natur.447..829P}.

\subsection{Constraints on II-L SNe progenitors}

There are very few direct constraints on nearby II-L SNe. This 
subtype appear to be relatively infrequent (see Table\,\ref{table1})
but they may be important in solving the problem of the 
lack of high mass red supergiants detected as type II-P progenitors. 
As the II-L by definition have a very short, or non-existent 
plateau phase they probably have a low mass H-envelope
which cannot sustain a lengthy recombination phase. 
The H-envelope mass could be reduced through mass-loss or 
binary mass-transfer. If the former, it could point to them 
being higher mass progenitors than II-P. 

The nearest II-L known, SN1980K  in NGC6946 (5.9\,Mpc) had a photographic
plate taken 49 days before maximum \citep{1982ApJ...257L..63T}. 
At the position of the SN there is no star, or stellar association 
visible to a plate magnitude of $M_{F} \simeq -7.7^{m}$. 
The limit does
rule out massive red supergiants greater than about 20\msol, but
 blue progenitors hotter than 10,000K and between 15-25\msol\
would be permitted. Another nearby type II-L SN1979C fell within 
a stellar association in M100 and analysis of the stellar 
population would suggest that if all stars were coeval the turn-off 
mass for the SN1979C progenitor would be 15-21\msol\ \citep{1999PASP..111..313V}. 
\cite{2000ApJ...532.1124M} have estimated the mass-loss history from the 
SN and find an increased rate at 10,000-15,000yrs before explosion. The 
total mass loss could be as high as 4-6\msol\ but they suggest this is 
not inconsistent with the stellar population mass. 
Absence of evidence is by no means evidence of absence, but 
to date there are  
no arguments from direct progenitor studies for high masses for II-L progenitors. 

\subsection{Are LBVs direct SNe progenitors ?}

\label{subsect:LBV-SNe}
The discovery of several remarkably bright, hydrogen rich (hence type
II) SNe has reinvigorated the debate of the physical mechanisms that
can produce explosions.  The first of these ultra-bright type II SNe
recognised was SN2006gy, followed by SN2005ap, SN2008es and
SN2006tf. The integrated radiated energies are around $10^{51}$\,ergs
and the physical cause of the exceptional luminosity is not yet established.  
The total energy of these explosions has not yet been  measured as the ejecta masses are uncertain, but typical kinetic energies of type II SNe also tend to be of order $10^{51}$ ergs. In
the case of SN2006gy and 2006tf (IIn SNe), 
\cite{2007ApJ...666.1116S,2008arXiv0804.0042S} propose that the
luminosity results from a physically similar process to that which
produces II-P SNe lightcurves (as discussed in Section\,\ref{sect:IIP})
but with extreme values for radial extent and density.  The shock
kinetic energy is thermalised in an opaque, dense shell (which acts
like a photosphere) of radius $\sim$150AU and mass of
$\sim10-20$\msol\ \cite{2007ApJ...671L..17S}. 
The radius and enclosed mass are too large to be a
bound stellar envelope, even when compared to the most extreme red
supergiants. Thus \cite{2008arXiv0804.0042S} propose that such dense
shells were created in LBV-like giant eruptions and mass ejections, within
a few years (perhaps up to decades) before final explosion.  In this model, 
the 
progenitor is required to be a massive LBV, one which is massive
enough to have undergone giant outbursts and by implication probably
greater than 50\msol. \cite{2008arXiv0810.0635A} developed a model
in which interaction is the luminosity source,  with an ejecta
mass of 5-15\msol\ impacting 6-10\msol\ of
opaque clumps of previously ejected material. Again this suggests
an LBV-type progenitor object.

The other two ultra-bright type II SNe (more correctly classed II-L as
they show no narrow absorption or emission components) SN2008es and
SN2005ap are equally luminous, again with total radiated energies
$\gtrsim10^{51}$\,ergs \citep{2007ApJ...668L..99Q,2008arXiv0808.2193M}. 
\cite{2008arXiv0808.2812G} offer an
alternative explanation for SN2008es of a progenitor with a lower mass, extended
H-rich envelope ($R\sim6000$\rsol) having a steady, dense super-wind with
mass-loss rate $\dot{M} \sim 10^{-3}$\msol\,yr$^{-1}$. 
For SN2005ap \cite{2007ApJ...668L..99Q}  suggest the 
collision shock and thermalization and also the possibility of a 
jet explosion (GRB-like) within a H-rich massive progenitor. 

Lightcurves powered by radioactive decay of  $^{56}$Ni 
were also considered \citep{2007ApJ...666.1116S,2008arXiv0808.2812G}
but this requires a huge mass of  $^{56}$Ni  in the ejecta ($\sim$20\msol). 
The sharp decline in the 
late-time lightcurves and lack of strong [Fe\,{\sc ii}]  lines
now suggests this is unlikely. Such a large $^{56}$Ni
mass could only be produced in a pair-instability supernova in 
which the high temperatures in a massive core (He cores of 
$\gtrsim$40\msol) induces electron-positron pair production. This
absorbs thermal energy, the core collapses further which results
in a further  temperature rise and runaway thermonuclear burning in a 
massive core \cite[for the details of the physics involved and review of 
the history of this idea see]{1986ARA&A..24..205W,2002RvMP...74.1015W}. 
In theory 10-20\msol\ of $^{56}$Ni can 
be produced and ejected \citep{2002ApJ...567..532H} in a pair-instability
supernova or $\sim$5\msol\ in a core-collapse of a massive star
\citep{2008ApJ...673.1014U}. 
A modification of this 
mechanism is pulsational pair-instability in which a massive
core undergoes interior instability again due to electron-positron
pair production
\citep{2007Natur.450..390W}. This leads to an explosion which ejects 
several solar masses of material, but is not enough to
unbind the star. Several pulsational explosions can occur
and the collisions between the shells could conceivably
produce $10^{50}$\,ergs. Again, the shock 
kinetic energy diffuses thermally within an optically
thick, high density, compact sphere. This produces the 
high luminosity rather than it being due
to a large mass of $^{56}$Ni. The model of  
\cite{2007Natur.450..390W} requires a large core mass 
from a star of initial mass 95-130\msol. 
The collisions between the massive shells produces radiative energies in a similar manner to that discussed in \cite{2007ApJ...671L..17S}

The radio lightcurve modulations seen in some SNe 
have been suggested to be due to the interaction of the ejecta
with the progenitor stars' surrounding gas shells which were
ejected in S-Doradus type variability \citep{2006A&A...460L...5K}. 
This would point to  stars which had been in the LBV phase
close to the epoch of  collapse. 
Additionally a direct LBV progenitor was also proposed for SN2005gj to 
explain the multiple components in the absorption trough of
H$\alpha$ \citep{2008A&A...483L..47T}. 

The physical mechanism that produces the ultra-bright type IIn and 
II-L SNe is still controversial and unresolved. Viable explanations
are the explosion of the most massive stars we know, while they
still retain a significant H-rich envelope or have recently 
undergone large mass ejections. Such objects are clearly
reminiscent of known LBVs in the Local Group. These massive 
stars are  in a position of the HRD that leads stellar evolutionary tracks
to suggest they are at the end of core H-burning or perhaps
have just entered core He-burning. If they are in fact 
undergoing core-collapse then their cores are significantly
more evolved than we have thought. This would pose 
difficulties for stellar evolution models and our 
interpretation of the nature of known LBVs. It is also not 
yet understood if the core-collapse mechanism (i.e. collapse
of an Fe-core and neutrino driven explosion)  can account for 
the energies observed. 

\newpage
\section{Explosion parameters and compact remnants}

The physics that governs the core-collapse and launch of the
shock that destroys the star has been of interest since 
the luminosities of SNe were first estimated. The current view
is that the shock bounce of the proto-neutron star requires
reinvigorating and boosting by neutrino energy deposition. 
\citep{2007PhR...442...38J}. 
Successful explosions have been produced numerically, 
but within restricted mass ranges. Acoustic wave 
driven explosions have also been proposed to increase the 
shock energy \citep{2006ApJ...640..878B}. 
The observations of progenitors do not give
restrictive constraints on the mechanisms by themselves 
but by comparing with the explosion parameters they are 
of interest to the core-collapse mechanism.

\subsection{$^{56}$Ni production and explosion energies}
\label{subsect:56Ni}
One of the few direct observational 
probes of the explosion which can be studied after core-collapse 
is measuring the amount of radioactive $^{56}$Ni that is
synthesised. This nuclide is created by the explosive burning of 
Si and O as the shock wave heats the surrounding mantle and is mixed
through the ejecta. The lightcurves of type Ibc and Ia SNe
around peak 
are determined by the mass of $^{56}$Ni, the total mass of the ejecta
and its kinetic energy \citep{2000ARA&A..38..191H,2006ApJ...645.1323M,2008MNRAS.383.1485V}. 
Models of the observed lightcurves and spectral 
evolution of Ic SNe have derived these properties 
\cite[e.g.][]{2006ApJ...645.1323M,2006NuPhA.777..424N,2008IAUS..250..463N}

The photospheric stage of II-P SNe is powered by the recombination of 
hydrogen as the photosphere cools but the nebular tail phase 
luminosity is determined by the $^{56}$Co$\rightarrow$$^{56}$Fe decay and its
subsequent deposition of $\gamma$-rays and positrons which are 
thermalised. Thus the bolometric luminosity in the nebular phase of type 
II SNe can be used to estimate the original $^{56}$Ni mass. 
There is a large range in the observed tail phase luminosities of 
type II-P SNe (e.g. see Figure\,\ref{fig:II-P-BLCs}) and the physical interpretation has been 
differences in the ejected $^{56}$Ni mass (for reference, the $^{56}$Ni 
mass estimated for SN1987A is 0.075\msol). 
\cite{2003MNRAS.338..711Z} and \cite{2004MNRAS.347...74P,2006MNRAS.370.1752P}
have measured  masses of $^{56}$Ni a factor of 10 lower (than for SN1987A)  in 
1997D, 1999br, 2005cs. These SNe also show low luminosity plateau
magnitudes, low ejecta velocities and hence low kinetic energies. 
The interpretation of \cite{2006NuPhA.777..424N}, 
\cite{2003MNRAS.338..711Z} and \cite{2004MNRAS.347...74P} is that they are 
initially high mass stars which result in faint explosions
(see Figure\,\ref{fig:Nimass}a). 

However the initial masses are dependent on the lightcurve model and
at least for some faint type II-P SNe there are direct 
progenitor mass estimates (Figure\,\ref{fig:Nimass}b). For  these  
 there is
no evidence of a massive progenitor, 
which  allows no confirmation of the massive 
progenitor and black-hole forming scenario. However there is 
still a possibility of there being two populations of 
faint SNe - one from massive progenitors as the lightcurve 
models and ejecta masses of 
\cite{2003MNRAS.338..711Z} and \cite{2006NuPhA.777..424N}
propose and one from the lower mass stars. This should
be testable as time allows larger numbers of progenitors to be
detected and the SN energetics quantified. In fact it should
be relatively easy to detect the high mass progenitors. 
If they are around 20-30\msol\
then they should have $ -8 < M_{\rm bol} < -9$, which are 
easily detectable in the images of the quality discussed in 
Sections \ref{sec:extragstellar} \& \ref{sect:IIP}. 
In Figure 11 the lack of a high luminosity branch in the nearby 
SNe with progenitor information is probably a selection effect as
these SNe are intrinsically rare and we have not had the 
opportunity to search for progenitors of their nearby analogues.

As it stands, the masses from direct detections and limits for
progenitors suggests there is an order of magnitude scatter in
the mass of $^{56}$Ni created in the explosions of stars
of seemingly similar masses. This is not well understood
within the current paradigms of  stellar evolution or explosion physics. Weak 
explosions from electron capture SNe have been proposed 
\citep{2006A&A...450..345K}
but these occur after 2nd dredge up when the progenitors
would be S-AGB stars and hence rather luminous, 
\logl$\simeq10^{5}$
\citep{2004MNRAS.353...87E,2008ApJ...675..614P}. 
\cite{2007MNRAS.376L..52E} show that SN2005cs for example was unlikely to have
been a S-AGB star. The diversity in explosion properties
of stars with apparently similar progenitor masses 
could reflect dependence on the exact density 
profile above the core, the rotation rate, chemical 
composition, or stellar magnetic field. As discussed by 
many modellers 
\citep[e.g.][]{1986ARA&A..24..205W,1987ApJ...322..206N,2002RvMP...74.1015W,2004MNRAS.353...87E}
the computation of evolution, and subsequent explosion, of 8-11\msol\ stars is 
complex due to electron degeneracy phases, thermal pulses and dredge-up. 

An example of further diversity in the explosions of
stars of probably similar mass is shown in Figure\,\ref{fig:II-P-BLCs}. 
In this case the bolometric lightcurves of the 
well studied SN1999em, SN2004et and SN2005cs and 
SN2003gd are compared. 
The distance to each galaxy is relatively well known and the
monitored flux covers from the UV to the NIR in each case. 
The progenitors have masses between 8-15\msol\ and are likely
red supergiants. There appears to be little correlation of 
kinetic energy, $^{56}$Ni mass or plateau luminosity 
with
progenitor mass. The progenitors of SNe 2005cs and 2003gd appear
very similar but their $^{56}Ni$ mass and kinetic energies 
differ by a factor of around 5. 
SN2003gd has a similar kinetic energy to SN1999em but their 
tail phase luminosity are significantly different with the inferred
$^{56}$Ni mass a factor of 3 lower in the case of SN2003gd.
This large diversity of explosion parameters from apparently
quite similar progenitors is puzzling. It will be of great
interest to see how the energy and luminosity of SN2008bk
compares as it was another explosion of a fairly low mass red supergiant 
(Section\,\ref{subsect:08bk}). 

The differences between the observed characteristics of II-P 
SNe in particular has previously been attributed to large
differences in the progenitor mass and radii 
\citep{2003ApJ...582..905H,2003MNRAS.346...97N,2008arXiv0809.3766U}. 
However the ejecta masses have not given good agreement with the 
direct masses of progenitor stars. Future work to reconcile the 
hydrodynamic ejecta masses and stellar evolutionary masses, which will
help quantify the explosion energies better is highly desirable.

\newpage
\subsection{NS and magnetar progenitors : turn-off masses}

\cite{2005ApJ...622L..49F} suggest that the soft gamma repeater
SGR 1806-20 lies within a stellar cluster with an age of
$\sim3-4.5$\,Myr. Assuming that the progenitor was coeval with the
starformation episode that created the cluster this would imply 
a mass of greater than $\sim$50\msol. SGRs are thought to 
be magnetars, which are slowly rotating ($P\sim$1-10 sec)
highly magnetized ($B\sim10^{14}$G) neutron stars. 
\cite{2000ApJ...533L..17V} suggest that SGR1900+14 was
born within a dense stellar cluster. An age estimate of the 
stellar population has been prohibitively difficult
due to difficulties in identifying a main-sequence turn-off. However the 
two M5 supergiants have bolometric luminosities 
which might suggest masses of between 8-12\msol\ assuming
the largest distance of 15\,kpc 
\cite[based on the RSG parameters of][]{2005ApJ...628..973L}.

\cite{2006ApJ...636L..41M} have discovered an x-ray pulsar 
only 1.7 arcmin from the core of the massive, young cluster
Westerlund 1. The age from the most massive stars in the cluster 
is $4\pm1$\,Myrs suggesting a progenitor mass for the X-ray 
pulsar of $>40$\msol, if it is associated and coeval. The 
x-ray luminosity and slow rotation period are more consistent with 
it being a 
magnetar.

\cite{2008ApJ...683L.155M} further suggest that the $\gamma$-ray 
source HESS J1813-178 may be part of a coeval association which 
includes two SN remnants and a cluster of massive stars with
ages of 6-8\,Myrs. This would imply a minimum mass of 20-30\msol\
for the progenitor. The likelihood of association between the
$\gamma$-ray source and the stellar population is the weakest of
these three and the nature of the high energy emission is not 
yet established. 

These four coincidences provide some evidence for very high mass progenitors
 of magnetars (40-50\msol), but this requires further investigation
as at least one example suggests a lower stellar population mass
and the association of HESS J1813-178 with a nearby stellar association
is not yet convincing. How neutron stars form from very massive progenitors is puzzling and further work in this area is imperative. 
\newpage

\newpage
\section{An overview and comparison with massive stellar populations}
\label{sect:overview}

\subsection{The lower mass limit for core-collapse}
\label{subsect:WDSNe}

The lower mass limit to produce a SN through core-collapse has
theoretically been suggested to lie between 7-11\msol. The mass estimates and
limits from Section\,\ref{sect:IIP} (see Figure\,\ref{fig:IIP-HRD})
for the II-P SNe provide a minimum mass estimate of
$m_{min}=8.5^{+1}_{-1.5}$\msol\ and  this can be taken as an
observational estimate for the minimum mass that can produce a
core-collapse.  The maximum stellar mass that produces white dwarfs in
young stellar clusters has been estimated to be no less than
$6.3-7.1$\msol\ at 95\% confidence by \cite{williams08,2008AJ....135.2163R}. 
It is not known if the
most massive white dwarfs (1-1.2\msol) have CO or ONe cores. 
Combining this with the fact that three RSG progenitors of II-P SNe
have been unambiguously detected with very similar estimated
masses (7-9\msol; Figures\,\ref{fig:IIP-images} and \ref{fig:IIP-HRD})
would suggest a convergence toward $8\pm1$\msol for the lower limit
to produce a SN. 
It should be noted that the WD masses and the RSG progenitor masses both depend on stellar evolutionary models and also WD cooling tracks and the bolometric luminosity model for RSGs. 

The models of  
\cite{2008ApJ...675..614P} and others (see references therein) suggest that
in the range 7.5-9.25\msol\ 
they become Super-AGB stars (S-AGB) and form an oxygen-neon core 
\citep{1984ApJ...277..791N}. 
The most massive  ($9-9.25$\msol) can reach the Chandrasekhar limit and explode as
ECSNe (see Section\,\ref{subsect:56Ni})
while above 9.25\msol\ normal Fe core collapse
occurs.
The stellar models predict high luminosities for the 
S-AGB progenitors of \logl$\sim$5.0\,dex, significantly higher than
any of the progenitors observed and above most of the upper limits. 
\cite{2008ApJ...675..614P} suggest that only a few
($\sim$3\%) of SNe are likely to be ECSNe. We certainly do see
weak explosions with low ejecta masses of $^{56}$Ni (e.g. 
see Figure\,\ref{fig:Nimass}) but in the cases of 2005cs and 
2003gd the progenitor was not a luminous S-AGB star 
\citep{2007MNRAS.376L..52E}. It maybe that these were weak 
ECSNe as the $^{56}$Ni and explosion energies were similar to
those of the explosion models of \cite{2006A&A...450..345K}, 
but the stars did not undergo 2nd dredge-up to become luminous. 

As discussed in Section\,\ref{subsect:OTs}
the possibility remains that the transients SN2008S, NGC300-OT2008 and 
M85-OT2006 could be examples of ECSNe. 
\citep{2008arXiv0809.0510T} suggest that they might be relatively
common explosions and have gone undetected until recently. They also 
point out that the rarity of the stellar analogues 
in nearby galaxies would suggest the dust enshrouded phase is short. 
It remains to be seen if the rate and explosion energies of these
events are compatible with predicted SNe from SAGB star models.

\subsection{Comparison with Local Group massive stellar populations}

Within the Galaxy and the Local Group there is now a wealth of studies
of evolved massive stars, both hot and cool 
\citep{2003ARA&A..41...15M} 
and this population is a reasonable comparator sample to compare with the
SN progenitors we have discussed. 

The effective temperatures and bolometric luminosities of
Galactic and Magellanic Cloud RSGs have been revised with new 
model atmospheres\citep{2005ApJ...628..973L,2006ApJ...645.1102L}. 
Their inferred luminosities  have been substantially reduced so that
they appear up to \logl$\lesssim 5.6$ which corresponds to an initial 
evolutionary mass of 30\msol. It is likely that this is their
final resting place before explosion as the minimum initial 
mass for a star into a H-deficient WR star is 25-30\msol\ at 
around solar metallicity. \cite{2001AJ....121.1050M} 
studied the WR population in twelve Galactic clusters and 
show that at solar metallicity the
minimum initial mass to produce a WR through single star evolution
is above 25\msol.  This rises to above 30\msol\ in the LMC. 
\cite {2007ARA&A..45..177C} point out that there are few
Milky Way clusters apart from Westerlund 1, that  host both WRs
and RSGs. This implies that they come from quite separate 
progenitor mass ranges. Thus Local Group studies seem to 
have established, with some measure of confidence, 
that RSGs evolve from single stars 
with masses up to around 25-30\msol. At solar metallicity
it is likely that stars of 25\msol\ and above can form WN stars
(with more massive objects becoming WC stars) . At LMC metallicity
this initial mass for WR formation is 30\msol. Hence one would expect
RSGs in the range 8 to 25-30\msol\ to be viable progenitors for
type II-P SNe. Evolutionary models can reproduce this separation 
between the RSGs and WR stars by including suitable mass-loss rates
(see Figure\,\ref{fig:largeHRD} for example). 

\subsection{The red supergiant problem}

After just the first few years of intensive systematic 
searching for progenitors the lack of easy detection of moderately
massive and very massive stars became an interesting issue 
\citep{2003MNRAS.343..735S}. The compilation of progenitor 
masses produced by \cite{2007ApJ...661.1013L} showed an obvious trend
and lack of high mass stars.   The volume and time limited survey
of \cite{SECM08} allows a statistical analysis of the mass ranges that
produce type II SNe and type II-P in particular.  As discussed in
Section\,\ref{sect:IIP}, the 20 II-P SN progenitors can be adequately fit with a
Salpeter IMF, a minimum mass of $m_{min}=8.5^{+1}_{-1.5}$\msol\ and a
maximum mass of $m_{max}=16.5\pm1.5$\msol. Comparing this to the 
Local Group massive stellar populations immediately raises the
question of the lack of detected RSG progenitors with initial 
masses between 17-30\msol. \cite{SECM08} term this the ``red supergiant
problem''. There are a number of possible explanations:

\begin{itemize}
\item The galaxy integrated IMF of massive stars could be significantly 
steeper than $\gamma = -2$. It would need to be at least $\gamma = -3$
to reduce the lack of massive RSGs to a statistically
insignificant number. \citep{2006MNRAS.365.1333W}
argue that  galaxy integrated IMFs could be steeper than Salpeter
due to the maximum stellar mass being linked to its natal cluster
mass. 
\item All massive stars above 17\msol\ could produce IL-L, IIn and Ibc SNe. The relative
frequencies of the II-P SNe compared to all other core-collapse types
match the stellar numbers from an IMF between 8.5-17\msol. For this 
to happen the II-L and IIn SNe must play an important role
which would mean severe mass loss occurs during the last stages 
of evolution of all massive stars.
\item Related to this, perhaps the metallicities of the progenitor stars
have been underestimated. If mass-loss rates can be extrapolated 
to higher metallicities than solar (and there is no evidence at present
that they can be) then perhaps WR stars can be produced from lower
masses than currently estimated at solar to LMC metallicity. 
\item Perhaps massive RSGs undergo severe mass-loss during the last
1-5\% of their lifetimes and become obscured in a dusty 
envelope which is optically thick at visible and NIR wavelengths
\cite[dusty red supergiants are known in the LMC;][]{2005A&A...438..273V}. 
Hence
the detections and limits reviewed in Section\,\ref{sect:IIP}
could be biased against these
stars, although the explosion would need to fully destroy the dust
envelop as the SNe themselves do not appear extincted. 
\item The massive RSGs that are visible in the Local Group between 
\logl$=4.0-5.5$\,dex and $M_{initial}\sim17-30$\msol\ do end their lives
in this evolutionary phase. But they produce SNe so faint that they 
have not been detected yet. An explanation for this is that their
cores form black-holes with no, or extremely weak, explosions
\citep{1999ApJ...522..413F,2003ApJ...591..288H}. 
\end{itemize}

If any one of these five explanations is the main reason then 
it has important implications for both SN studies and massive
stellar evolution. If a steep, galaxy integrated IMF is the cause
it would have far reaching implications \citep{2006MNRAS.365.1333W}. 
One could imagine that it is a combination of the first four and
that we could stretch each of the current best estimates of the 
IMF, initial mass for WR formation, metallicity and metallicity 
dependent mass-loss and RSG extinctions by a reasonable amount
so that the cumulative effect could account for the observations. 
All the effects would need to conspire to work in unison however. 

\subsection{Mass ranges for progenitors}

The most intriguing possibility is that we are seeing the first
observational signals for the stellar mass range that form black-holes
in core collapse.  This is perhaps the explanation 
 that would cause least
contradiction with known parameters of massive stellar populations. 
Models have predicted that  between about \zsol\
and 0.5\zsol, stars with initial masses above 25\msol\ may not be
able to explode through the presumed core
bounce and neutrino driven mechanism. This might suggest that red
supergiants above 25\msol\ and massive WR stars from initial masses
above 30\msol\ collapse quietly to form black holes and 
either very faint SNe or none at all
\citep{1999ApJ...522..413F,2003ApJ...591..288H}. In Section\,\ref{sect:Ibc}
one could draw a conclusion from the review of the limits on Ibc SNe and the 
measured ejecta masses that all Ibc SNe (which are not
broad-lined or associated  with GRBs)
arise from interacting binaries from progenitors with initial
masses 8-15\msol. It could be that the more massive cores 
form black-holes and produce Ic SNe and GRBs through the collapsar 
mechanism. In this case the difference between quiescent collapse
and a jet induced explosion would be angular momentum of the CO 
star. This would mean virtually all (probably 95\% ; see Podsiadlowski
et al. 2004) local WR stars do not produce Ibc SNe.
At first thought this is perhaps surprising and controversial 
but this is not in serious conflict with any of the restrictive
observational studies of SNe progenitors reviewed here. 
The case of SN2008ax suggests that single WN stars (of initial mass
around 25\msol) can produce bright IIb SNe so there may not be a sharp
mass cut-off between the two types and it may be smeared due to 
other effects like metallicity, rotation and mass-loss. 
An interesting area for future work would be a survey for quietly disappearing
massive stars as suggested by \cite{2008ApJ...684.1336K}.

Attempts have been made in the past to extend the simple picture of
the ``Conti scenario'' of massive stellar evolution in which mass-loss
drives the schematic evolutionary phases of massive stars
\citep{1976MSRSL...9..193C}. Variations on such extensions were
discussed by \cite{2003ARA&A..41...15M, 2007ARA&A..45..177C}
and \cite{2007ApJ...656..372G} for example. However these are overly simplified
 when one considers the added effects that metallicity,
rotation and binarity can play. This is not a criticism of the
schemes, merely a statement that a one dimensional evolutionary route
which is based on observational evidence is probably not
sufficient. Theoretical stellar population studies can quantify the
different effects of binary fractions, rotational velocity
distributions and metallicity with parameterized values giving
fractions of the SN types and tree diagrams
\citep[e.g.][]{1992ApJ...391..246P}. 
Hence an attempt is made in Figure\,\ref{fig:summary} to show the paths
to core-collapse that match what has been presented in this review. 
It is meant to illustrate the diversity and complexity of 
phenomena that are observed as well as giving a likely path. 
I should stress that this is not meant to be definitive and there
will be inevitable adjustments to the diagram as time progresses 
(particularly with regard
the new types of transients) but it summarizes the results
reviewed here and the bulk of the local SN population. 
 One problem with the figure is that it does not adequately deal with metallicity effects and as \cite{2008AJ....135.1136M} show, metallicity may play a critical role in defining the explosion mechanism and GRB production. 

\section{Summary points}

\begin{enumerate}
\item The progenitors of II-P SNe have been confirmed as red supergiants, 
although there has been a surprising lack of high mass stars detected. 
The three best detections still await confirmation that the progenitor
stars have indeed disappeared. The lack of high-mass progenitors
has interesting implications for stellar evolution and explosion mechanisms. 
The minimum mass that produces SNe seems to be converging toward
$8\pm1$\msol. 

\item It is almost certain that interacting binaries play an important 
role in influencing the relative rates of types within SN populations. 
The progenitor system of the SN1993J (a IIb SN) is well characterised 
and it appears very likely that a significant fraction of Ibc SNe come
from interacting binaries. 

\item There is a plausible candidate for a WR progenitor (probably
a WN star) of SN2008ax. This was a IIb hence indicating that 
different channels can produce similar, but not identical SNe. 
So far there is no confirmation that massive WR stars produce 
the majority of Ibc SNe in the local Universe. There are arguments supporting
them as progenitors of broad-lined, highly energetic Ic SNe which
are related to GRBs. 

\item Evidence now exists that LBVs or stars showing LBV like 
characteristics die in luminous explosions. The recent discoveries
of the brightest hydrogen-rich SNe known also suggests
high mass LBV type progenitors. The explosion mechanism which 
produces these is not easy to reconcile with an Fe-core collapse. 
New physical mechanisms are probably required. 

\item Three low-luminosity transients have been discovered which 
may have dust embedded massive star progenitors. Their nature
is currently uncertain but it is possible they are ECSNe
in Super-AGB stars. 

\end{enumerate}

\section{Future issues and prospects}

\begin{itemize}
\item Apart from extraordinarily bright progenitors from rare SNe, it
has been difficult to detect progenitors beyond about 10\,Mpc. Hence 
the greatest potential for future discovery in this field will come
from a concerted effort to gather
deep, multi-wavelength (from the UV to mid-IR) wide-field
imaging of nearby galaxies for future SN progenitor characterisation. 
This can be a combination of space and ground-based images. The SNe 
themselves require rapid and intense follow-up to characterise their
explosions. 

\item The new transients discovered at the extrema of the SN spectrum
(low and high luminosity) require further physical understanding. 
It may be that the canonical Fe-core collapse mechanism is
unable to explain the full range of explosion parameters and 
alternative explosion physics is required. This is an 
area ripe for intense theoretical and observational effort.

\item The rare ultra-bright events, intrinsically faint explosions and
SNe in low-luminosity metal poor hosts are likely to be discovered
in much larger numbers with future deep, wide-field optical 
surveys such as Pan-STARRS, SkyMAPPER, Palomar Transient Factory and 
eventually LSST. Potentially new types of stellar explosion
could be discovered by combining optical detections with 
LOFAR, Fermi, Advanced LIGO and neutrino experiments.

\item Exactly which type of stars produce stellar mass black-holes
is not yet understood and the lack of high mass progenitors 
may suggest there is a population of black-hole forming 
SNe which so far have eluded discovery. Searches for faint events, 
or perhaps no explosions at all are interesting areas for
future effort. 

\end{itemize}

\section*{Acknowledgment} 
It is a pleasure to thank the European Heads of Research Councils and
European Science Foundation EURYI (European Young Investigator) Awards
scheme, which was supported by funds from the Participating
Organisations of EURYI and the EC Sixth Framework Programme. 
I thank my current and former students and postdocs for their dedication, stimulating ideas, and article review, particularly Justyn Maund, Mark Crockett, Dave Young, Stefano Valenti, Andrea Pastorello, Seppo Mattila, Kate Maguire, and also to Robert Ryans for graphics assistance. John Beacom, John Eldridge, and Ken Nomoto,  Nathan Smith and Norbert Langer provided very useful comments on an initial draft. I thank Rolf Kudritkzi, Danny Lennon, and Gerry Gilmore for initially inspiring my interest in stars outside our Galaxy.

\section*{Key Terms and Acronyms} 
{\bf Acronyms} 

\noindent {\bf CCSN} Core collapse supernova \\

\noindent {\bf RSG} Red supergiant \\

\noindent {\bf BSG} Blue supergiant \\

\noindent {\bf IMF} Initial mass function \\

\noindent {\bf Bolometric lightcurves:}
Integrated flux from the UV to the infra-red 
usually 0.3-2.5$\mu$m, as a function of time,  
to monitor the total radiated energy. 

\noindent {\bf Type II-P SNe:}
SNe showing P-cygni H-lines and a long plateau in the lightcurve.
Expanding photosphere phase powered by recombination of hydrogen.

\noindent {\bf Type Ibc SNe:}
Classification into Ib or Ic categories can be ambiguous, Ibc is
often used as an umbrella term for both. 

\noindent {\bf Electron Capture core-collapse:}
A stellar core of ONeMg reaches the Chandrasekhar limit.  Electron
capture by $^{24}$Mg and $^{20}$Ne triggers collapse before O and Ne
are ignited

\noindent {\bf Luminous Blue Variables:} 
Massive luminous stars with H and He rich atmospheres and strong
winds. Variable photospheric temperatures and can undergo luminous 
outbursts. 

\noindent {\bf SN impostors:}
Some faint IIn SNe are actually giant eruptions of LBVs 
rather than core-collapse explosions - termed ``SN impostors''

\noindent {\bf Wolf Rayet stars:} 
Evolved massive stars that have lost their
envelopes through radiatively driven 
winds. They have high mass-loss rates, low He and H content
and are likely of original mass more than 25-30\msol\

\noindent {\bf WN} Nitrogen sequence Wolf-Rayet \\
\noindent {\bf WC} Carbon sequence Wolf-Rayet \\
\noindent {\bf WO} Oxygen sequence Wolf-Rayet 

\noindent {\bf Gamma ray bursts:}
Flashes of electromagnetic radiation with durations 
of order of seconds and photon energies $\sim$100\,keV. 
Isotropically distributed the vast majority are at
cosmological distances. 

\noindent {\bf Long duration GRBs:}
GRBs are broadly categorized into long-soft bursts 
(LGRBS ; typical duration $\sim$20s) and short-hard bursts
($\sim$0.3s). Total  $\gamma$-rays energy in LGRBs
is $\sim10^{51}$\,erg. 

\noindent {\bf Type Ic-BL:}
The nearest long duration bursts are coincident with highly energetic 
type Ic SNe - called ``broad-lined'' Ic or hypernovae. 

\noindent {\bf Ultra-bright type II SNe:}
A newly discovered group of SNe which have enormous
luminosities, typically $10^{51}$\,ergs integrated, 
$\sim$100 times more than normal CCSNe. 


\section*{Reference Annotations}

\noindent {\bf Crowther 2003:} Extensive review article on the physical parameters of 
massive WR stars. 

\noindent {\bf Gal-Yam \& Leonard 2009: } Discovery of a very luminous star,
probably an LBV, 
as  the progenitor of a IIn SN and evidence that it has since disappeared. 

\noindent {\bf Heger et al. 2003:} Theoretical models 
of stellar evolution are linked to the
type of SN and remnants produced as a function of metallicity. 

\noindent {\bf Massey 2003:} Review of the massive stellar populations 
in the Local Group. 

\noindent {\bf Pastorello et al. 2007:} First discovery of a luminous outburst 
before the collapse of a massive star and subsequent SN. 

\noindent {\bf Smartt et al. 2009:} Volume and time limited search for 
progenitors of II-P SNe, consistent analysis and statistical results for 
progenitor mass ranges. 

\noindent {\bf Smith et al. 2007:} First paper on the new class of 
ultra-bright type II SNe. 

\noindent {\bf Woosley \& Bloom 2006:} Review of the supernova - gamma ray
burst connection

\begin{table}
\caption{The relative frequency of core-collapse supernova types reported in 4 different studies. 
SECM \citep{SECM08}, LWVetal07 \citep{2007ApJ...661.1013L}, 
VLF08 \citep{2005PASP..117..773V}, 
PSB08 \citep{2008ApJ...673..999P}, CET99 \citep{1999A&A...351..459C}. The uncertainties are simple Poissonian 
errors and the total number of objects in each survey is listed in the {\em Sample Size} row. SECM08 and LWVetal07
are volume limited estimates with distance limits of 28\,Mpc and 30\,Mpc, covering different time periods. VLF05 
is based on LOSS discoveries within about 140\,Mpc. The PSB08 sample is between about $40-170$\,Mpc and 
CET99 combines various surveys mostly within 100\,Mpc.}
\label{table1}
\vspace{0.5cm}
\small
\begin{tabular}{rrrrrr}
\toprule
                      &         \multicolumn{5}{c}{Sample} \\  \hline

Type \vline         &  SECM08   \vline       &  LWVetal07     \vline   &  VLF05           \vline  &  PSB08        \vline  &  CET99          \vline    \\
\colrule
II-P         \vline  & 58.7$\pm$8.0\% \vline & 67.6$\pm$10\%   \vline  &  62.9$\pm$4.7\%  \vline  &75.5$\pm$9.8\%  \vline & 77.7$\pm$10.8\%  \vline  \\\cline{1-2}
II-L         \vline  &  2.7$\pm$1.7\% \vline &                 \vline  &                  \vline  &                \vline &                 \vline \\\cline{1-4}
IIn          \vline  &  3.8$\pm$2.0\% \vline & 4.4$\pm$2.5\%   \vline  &  9.2$\pm$1.8\%   \vline  &                \vline &                 \vline \\\cline{1-4}
IIb          \vline  &  5.4$\pm$2.7\% \vline & 1.5$\pm$1.5\%   \vline  &  3.2$\pm$1.0\%   \vline  &                \vline &                 \vline \\\hline
Ib           \vline  &  9.8$\pm$3.3\% \vline &                 \vline  &                  \vline  &                \vline &                 \vline \\\cline{1-2}
Ic           \vline  & 19.6$\pm$4.5\% \vline & 26.5$\pm$6.2\%  \vline  &  24.7$\pm$3.0\%  \vline  &24.6$\pm$5.6\%  \vline & 22.3$\pm$5.8\%  \vline \\\hline
Sample size  \vline  & 92             \vline & 68              \vline  &  277             \vline  &77              \vline & 67              \vline  \\
\botrule
\end{tabular}
\end{table}
\normalsize

\begin{figure}[ht]
\caption{\small{\bf (a)+(b)}: Colour image of the progenitor of SN2008bk. The
pre-explosion image  (a) is a combination of VLT optical and NIR images and the progenitor
is identified as a bright red point source. The Adaptive-Optics NACO $K_{\rm s}$ image
(near diffraction limited resolution of 0.1 arcsec) used for precise differential
astrometry was taken roughly two months after explosion. 
Both images are from \cite{2008arXiv0809.0206M}. \newline
{\bf (c)+(d)}: Colour image of the progenitor of SN2005cs. 
The pre-explosion HST ACS-WFC image (c) shows the red supergiant 
progenitor found to be coincident with SN2005cs by 
\cite{2005MNRAS.364L..33M} and \cite{2006ApJ...641.1060L}. The 
ACS-HRC image (d) shows SN2005cs as a bright blue source. 
These images of SN2005cs are archive data taken by Filippenko et al. 
(HST program GO10182; F330W images taken 46-50\,days after explosion)
and  Li et al. (SNAP 10877; F555W and F814W taken at 530\,days after explosion. 
\newline
{\bf (e)+(f)}: Colour composite showing the progenitor of SN2003gd using
the data presented in \cite{2004Sci...303..499S} and \cite{2003PASP..115.1289V} 
and supplemented with a late-time F450W archive image from SNAP10877. As the 
SN is not detected in that image, it can be used to construct the pre-explosion
colour composite.  The image of SN2003gd shown in (f) was taken 
from \cite{2004Sci...303..499S}, taken about
137 days after explosion. These examples show 
unambiguous RSG progenitors of three nearby type II-P SNe. \newline
Image Credit: David R. Young and R. Mark Crockett. 
}
\end{figure}

\begin{figure}[ht]
\centerline{\psfig{file=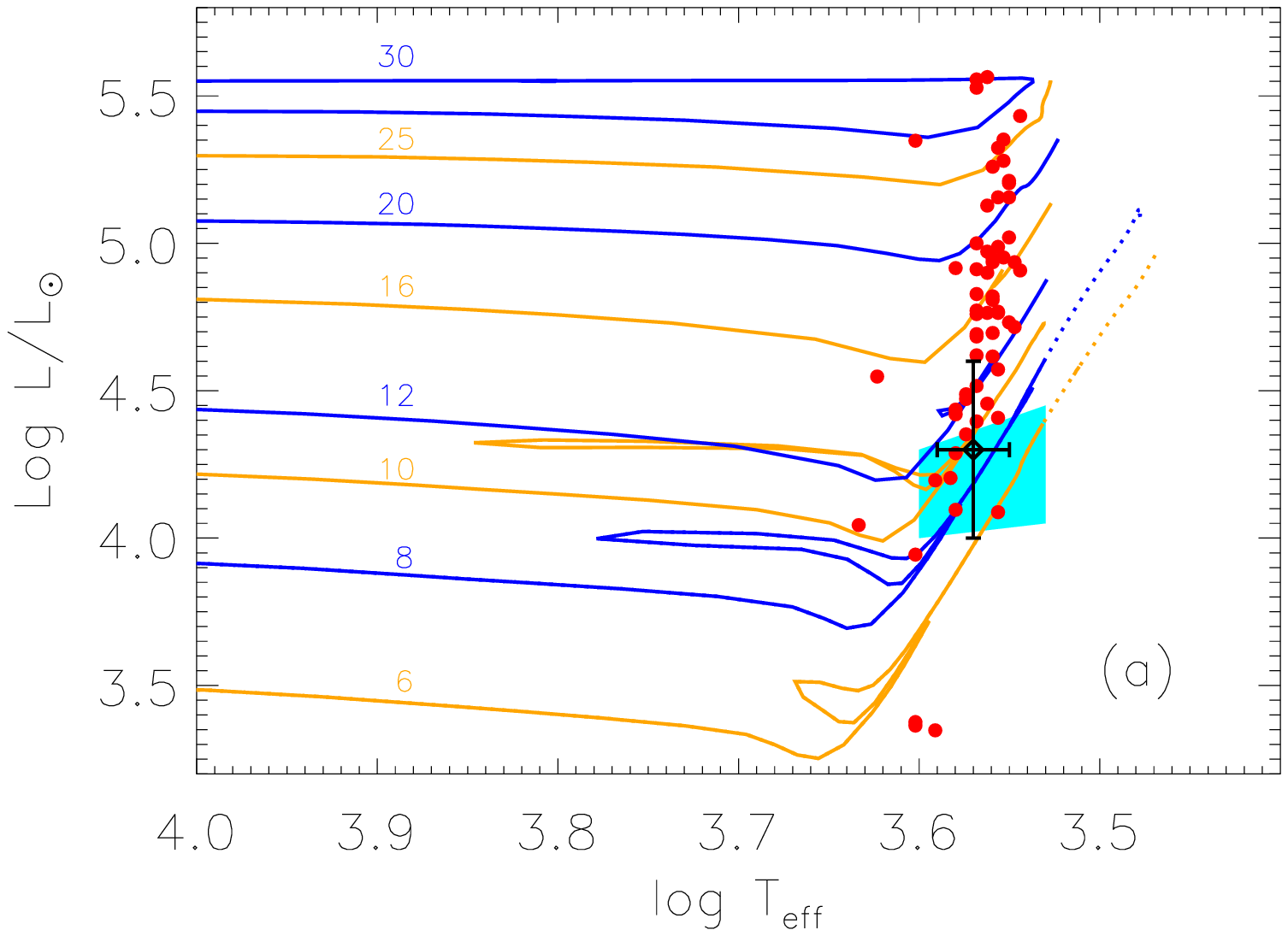,height=9cm,angle=0  }}
\centerline{\psfig{file=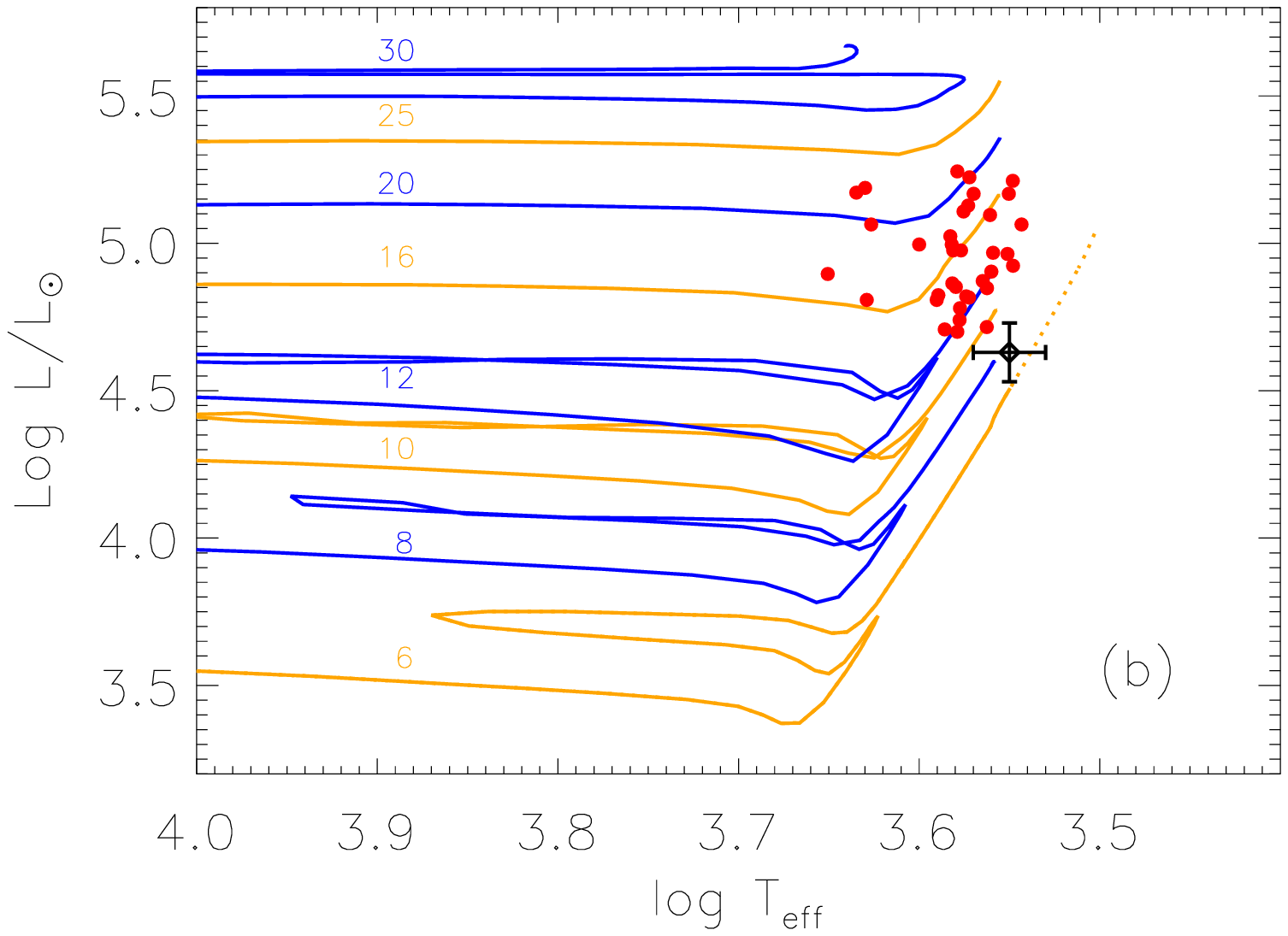,height=9cm,angle=0  }}
\caption{\small {\bf (a)} The progenitors of SN~2003gd (black error bar) and SN~2005cs (blue shaded
region with the STARS evolutionary tracks at $Z=0.02$ overplotted from masses 6-30\msol. 
The 6 and 8\msol\ tracks have the 2nd-dredge up phase indicated with 
the extended dotted track. The red points are the Milky Way red supergiants
from \cite{2005ApJ...628..973L}. {\bf (b):} The progenitor of SN2008bk 
with the LMC RSGs of \cite{2006ApJ...645.1102L} and the STARS tracks at $Z=0.008$. }
\label{fig:IIP-HRD}
\end{figure}

\begin{figure}[ht]
\centerline{\psfig{file=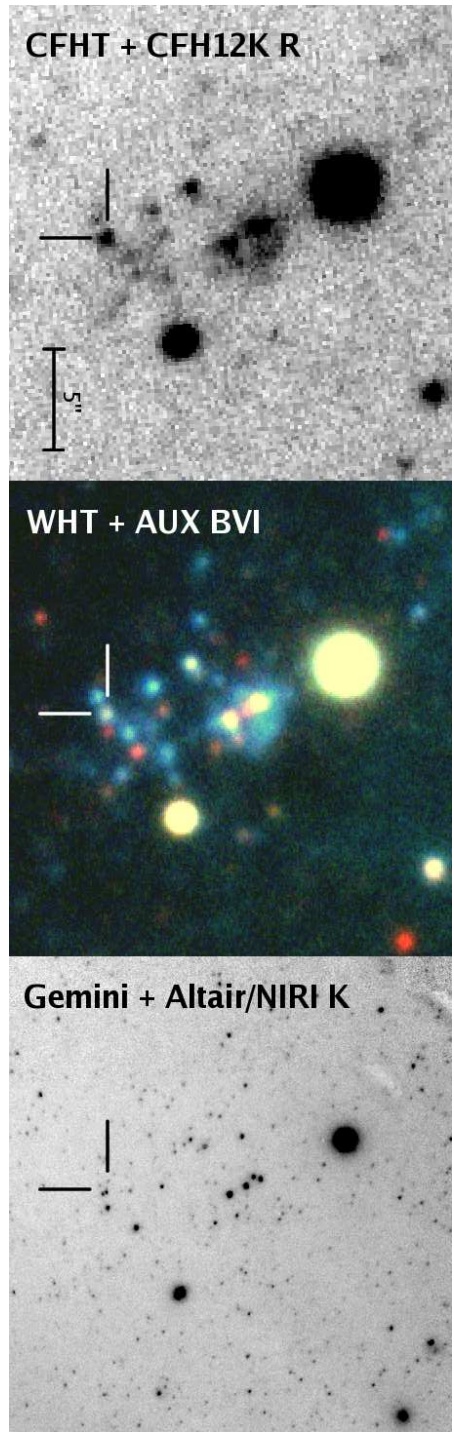,height=19cm,angle=270}}
\caption{\small  The progenitor of SN2004et was first proposed to be a
high mass 
yellow supergiant (Li et al. 2005), identified by the cross-hairs in the CFHT 
pre-explosion $R-$band image. However the WHT image (b) in the centre panel 4 years
after discovery shows the same source visible at the same $BVRI$ magnitudes. A
near diffraction limited $K-$band image from Gemini North clearly
reveals that the object identified as the progenitor of SN2004et was
not a yellow supergiant but a cluster of massive stars. The progenitor
originated within this small association and no evidence for yellow
supergiant progenitors now exist. Images from Crockett (2009). }
\label{fig:04et}
\end{figure}

\begin{figure}[ht]
\centerline{\psfig{file=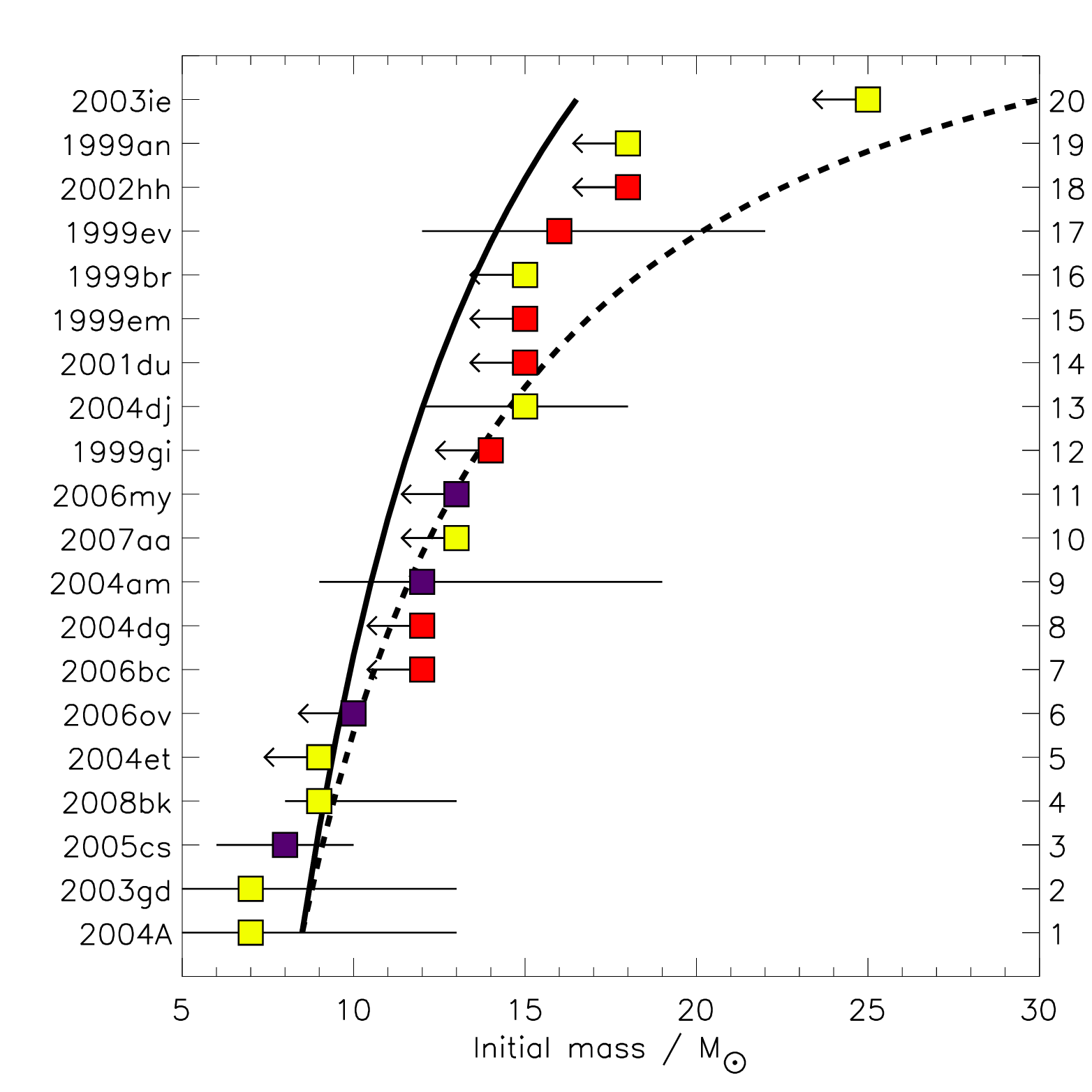,height=8cm,angle=0}}
\centerline{\psfig{file=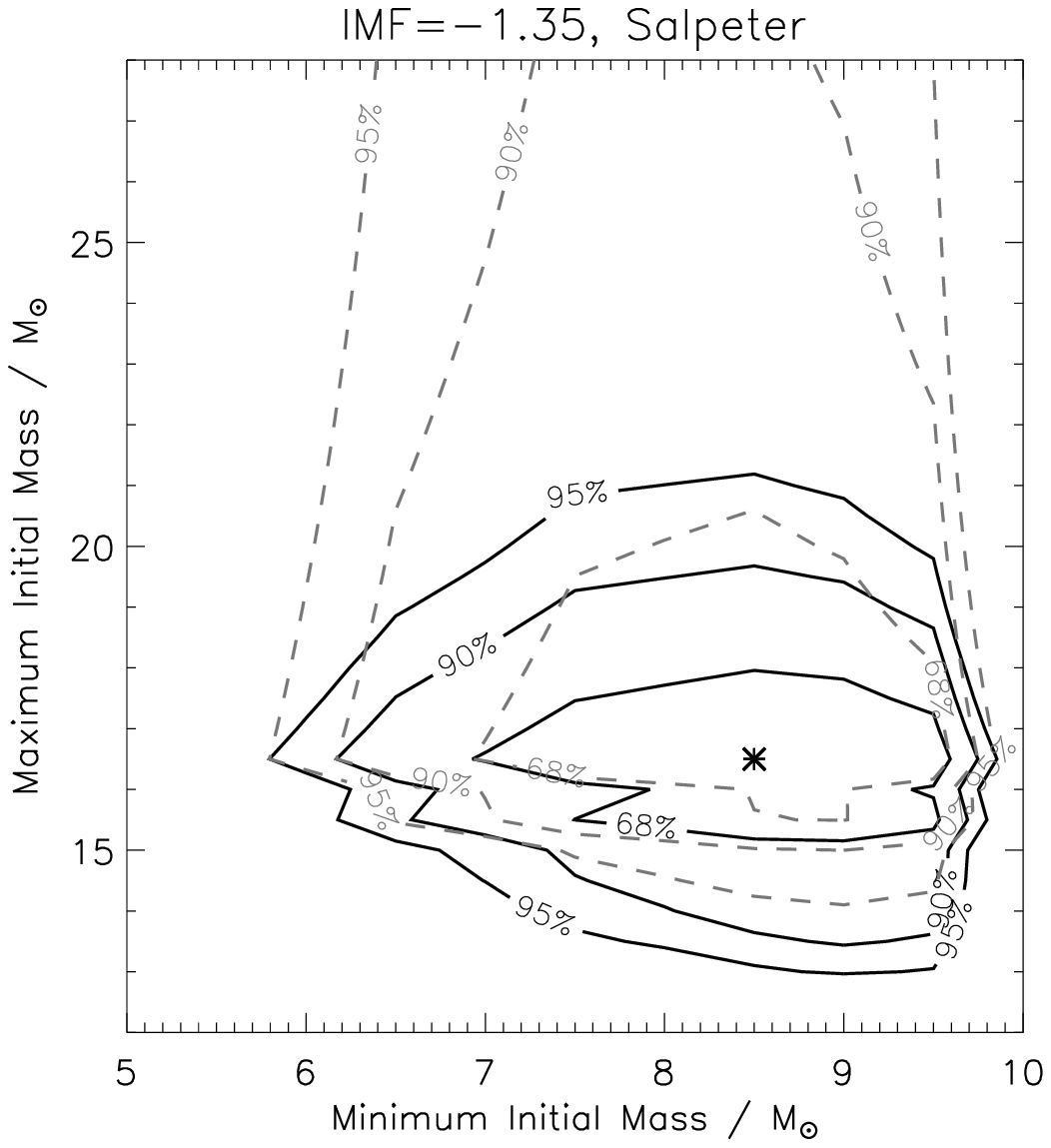,height=8cm,angle=0}}
\caption{\small {\bf (a):} A cumulative frequency plot of the masses of II-P progenitors, 
taken from Smartt et al. (2009). The right-hand axis is a simple number count
and the SNe are ordered in increasing mass or mass limit. The solid line
is a Salpeter IMF ($\alpha=-2.35$) with a minimum mass of 8.5\msol\ and
maximum mass of 16.5\msol\ which is the most likely fit to the data. The 
dotted line is a Salpeter IMF but with a maximum mass of 30\msol. The 
SNe are grouped in metallicity bins $\log {\rm O/H} + 12 = 8.3-8.4$ (yellow), 
8.5$-$8.6 (red), 8.7$-$8.9 (purple). 
{\bf (b):} The maximum likelihood analysis of the II-P progenitor sample
gives the most likely value for initial and final mass and the likelihood
contours \citep[also from][]{SECM08}. The dashed lines are those
calculated with detections only and the solid lines represent the 
contours calculated including the upper masses.
}
\label{fig:IIP-imf-max}
\end{figure}

\begin{figure}[ht]
\centerline{\psfig{file=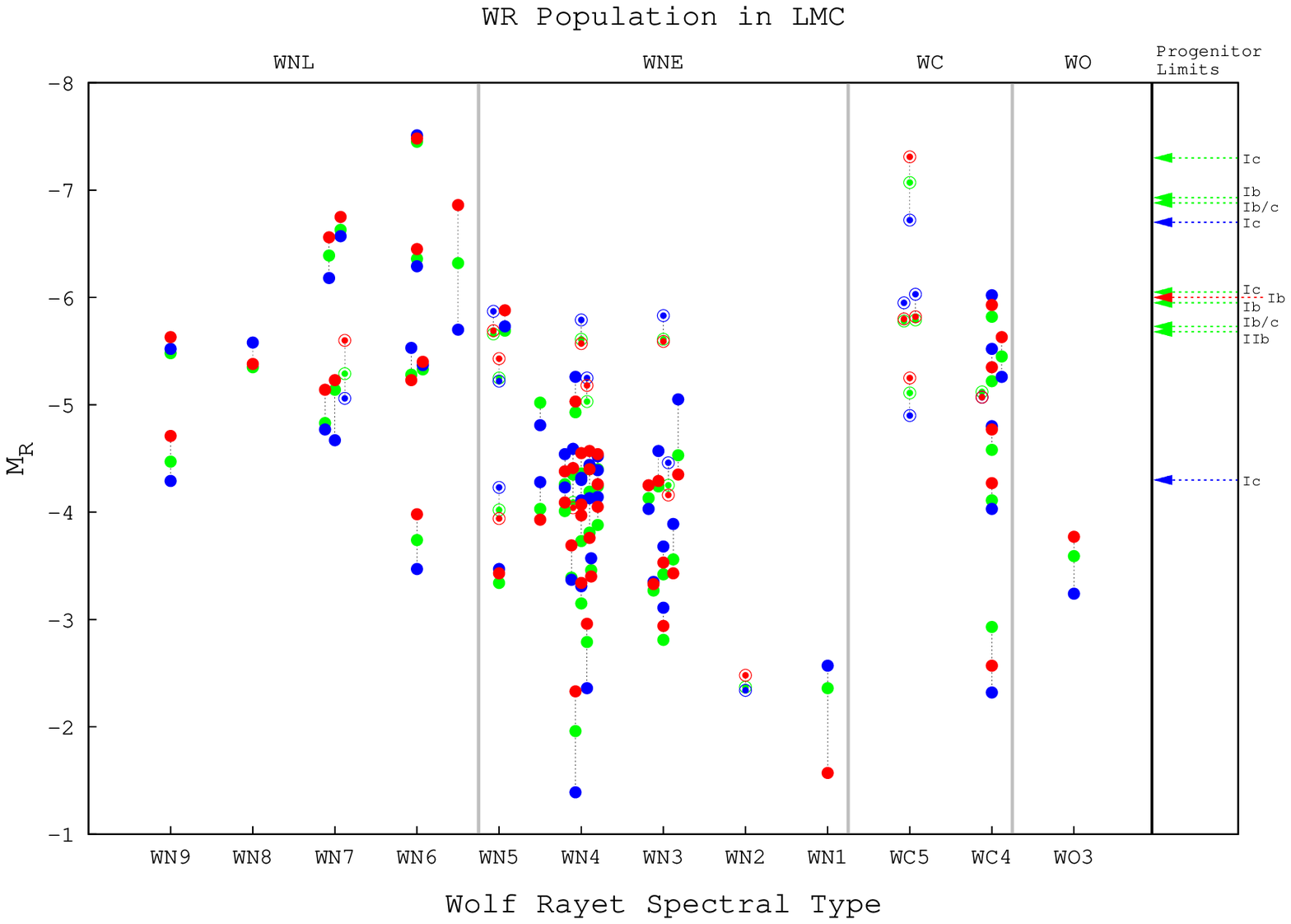,width=15cm,angle=0}}
\caption{The $BVR$ magnitudes (blue, green, red symbols) of WR stars in the LMC
(circled dots are likely binaries) from \cite{2002ApJS..141...81M} .
The magnitude limits for all Ibc SNe as discussed
in Section\,\ref{sect:Ibc} are shown on the right. If these massive stars are the 
progenitors of local Ibc SNe then there is only a 10\% chance we have not 
detected any by chance. The arrows are colour coded blue, green red to 
signify psuedo-$BVR$ limits respectively. Adapated from Crockett (2009). }
\label{fig:WRIbc}
\end{figure}

\begin{figure}[ht]
\centerline{\psfig{file=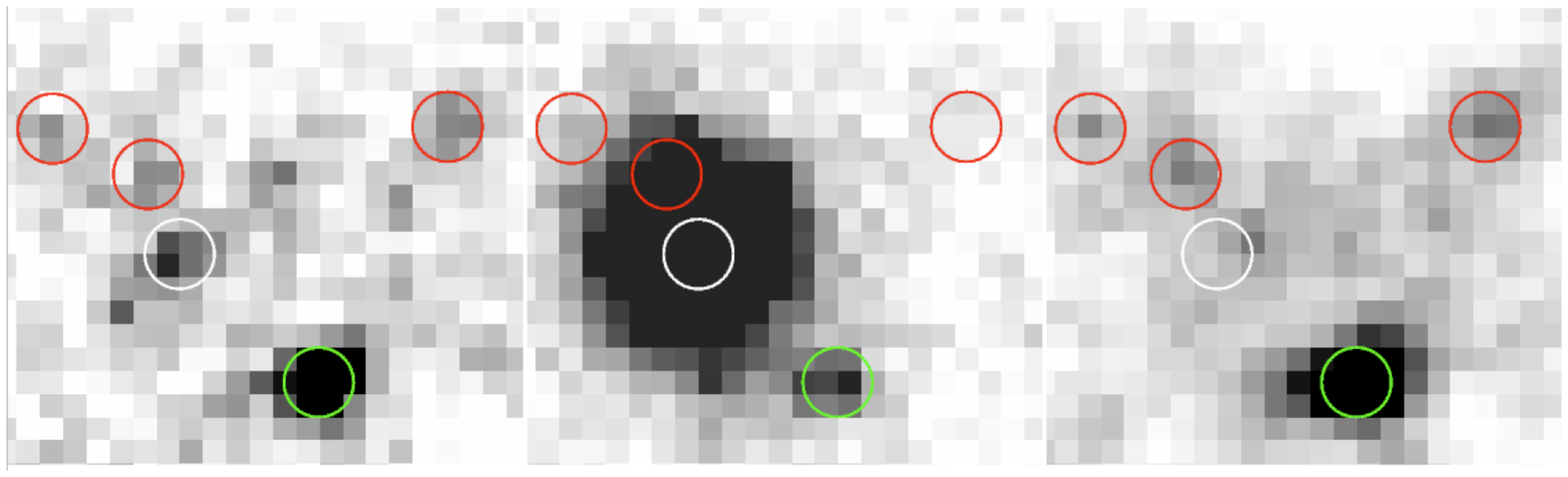,width=15cm,angle=0}}
\centerline{\psfig{file=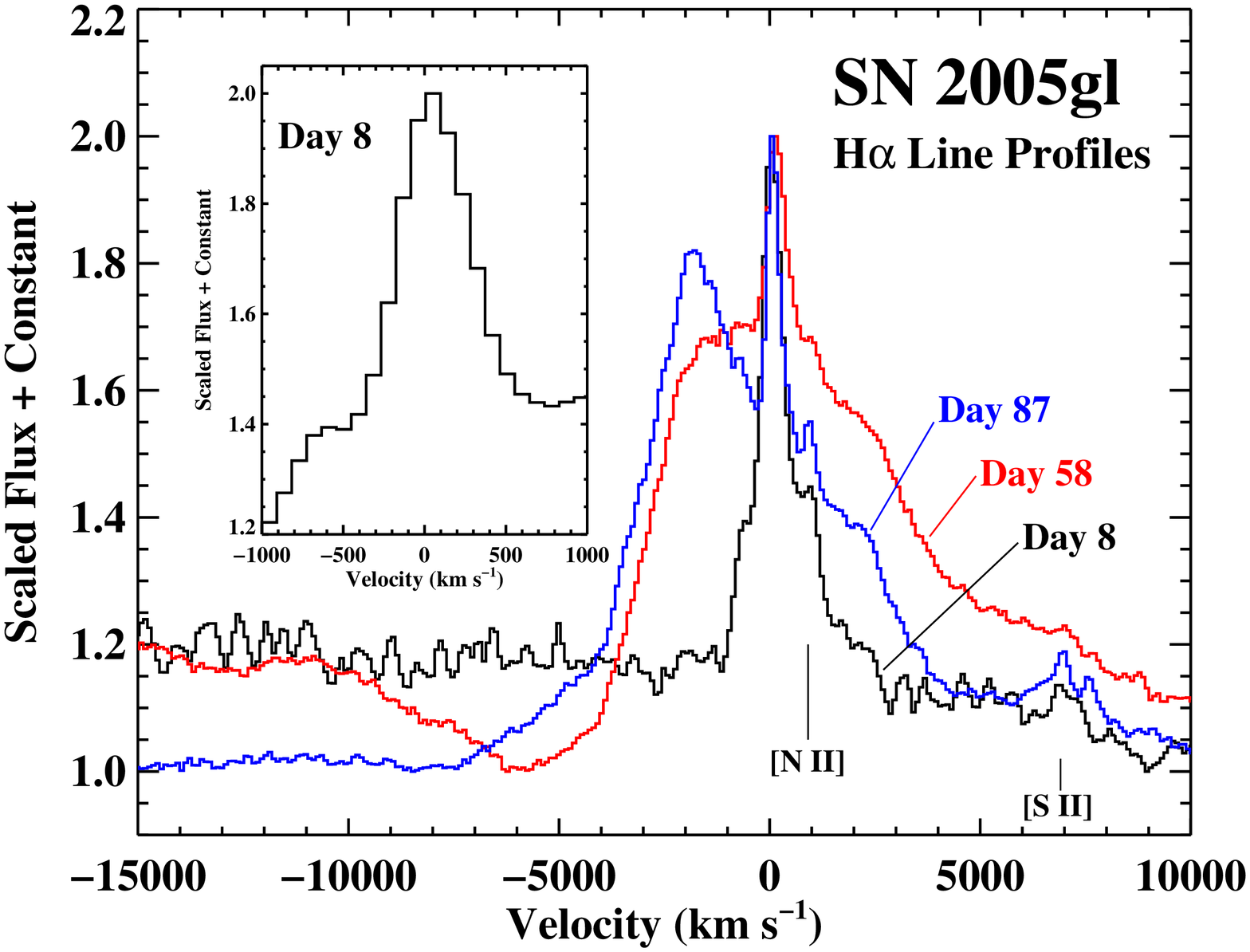,width=10cm,angle=90}}
\caption{\small The upper 
panels show the detection of the progenitor of SN2005gl
in a 1997 pre-discovery HST F547M image (within the white circle). 
The SN is shown in the middle panel from 2005 and is coincident
with the bright progenitor object from 1997. 
The repeat HST image taken in 2007 shows the progenitor star
has disappeared (again, position denoted by the white circle). 
The 
lower panel shows the evolution of the H$\alpha$ profile of SN2005gl, 
classified as a IIn. Early in the evolution, the profile is narrow 
suggesting excitation of a dense circumstellar medium and the broad eject
become visible later. All material is from \cite{gal-yam_leonard}. 
}
\label{fig:05gl}
\end{figure}

\begin{figure}[ht]
\centerline{\psfig{file=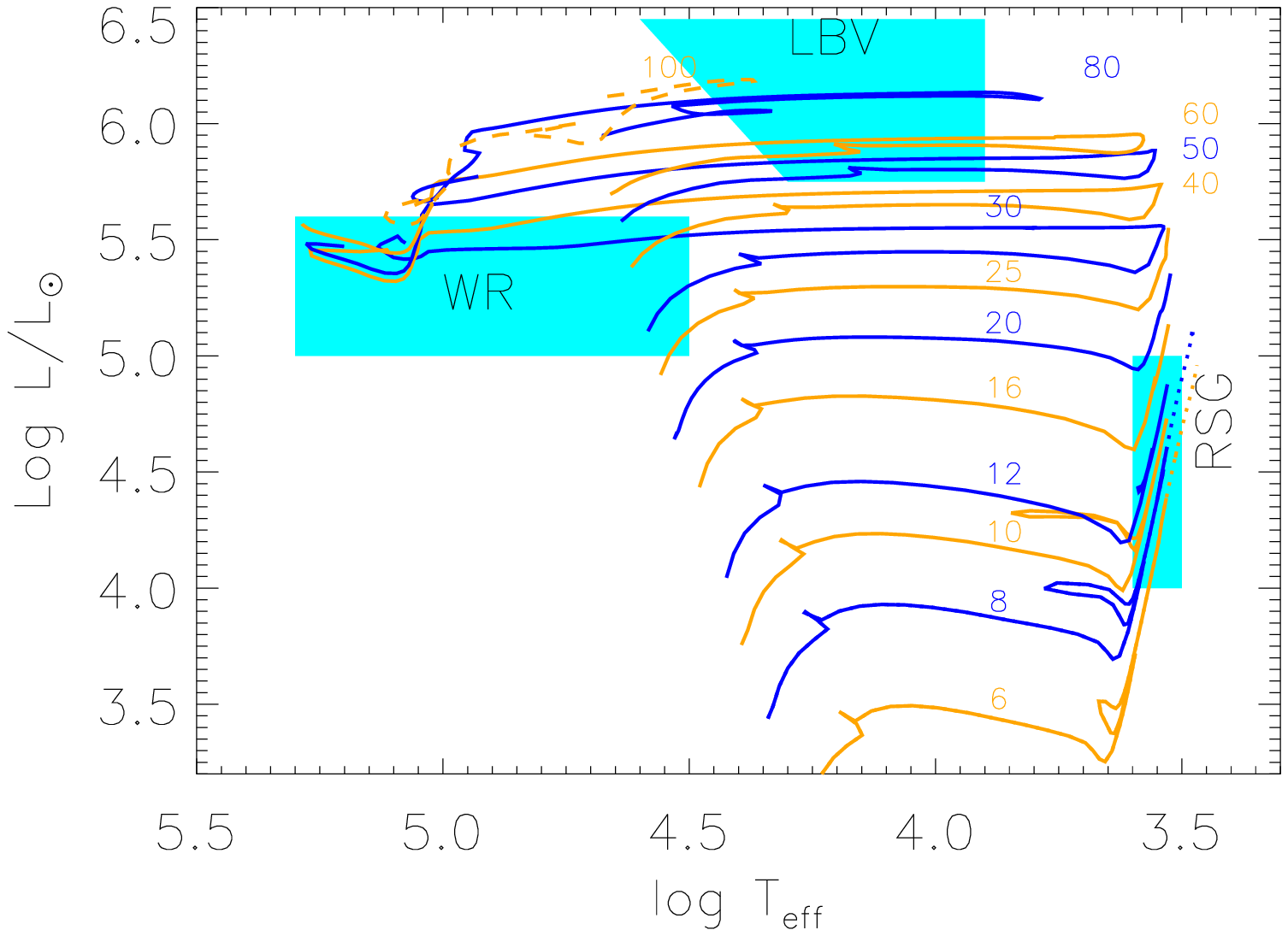,width=15cm,angle=0}}
\caption{The HRD of the STARS evolutionary tracks (Eldridge \& Tout 2004). The 
location of the classical LBV region from \cite{2004ApJ...615..475S} is illustrated. 
SN2005gl had a luminosity of at least \logl$\simeq10^6$, which puts it in the 
LBV region indicated, or at even higher luminosities if it was hotter and hence
had a significant bolometric correction. The region where we should see WR progenitors is 
shown and the only progenitor detected close to this region is that of SN2008ax. The
RSG region in which observed progenitors have been detected is shown again for 
reference.}
\label{fig:largeHRD}
\end{figure}

\newpage
\begin{figure}[ht]
\centerline{\psfig{file=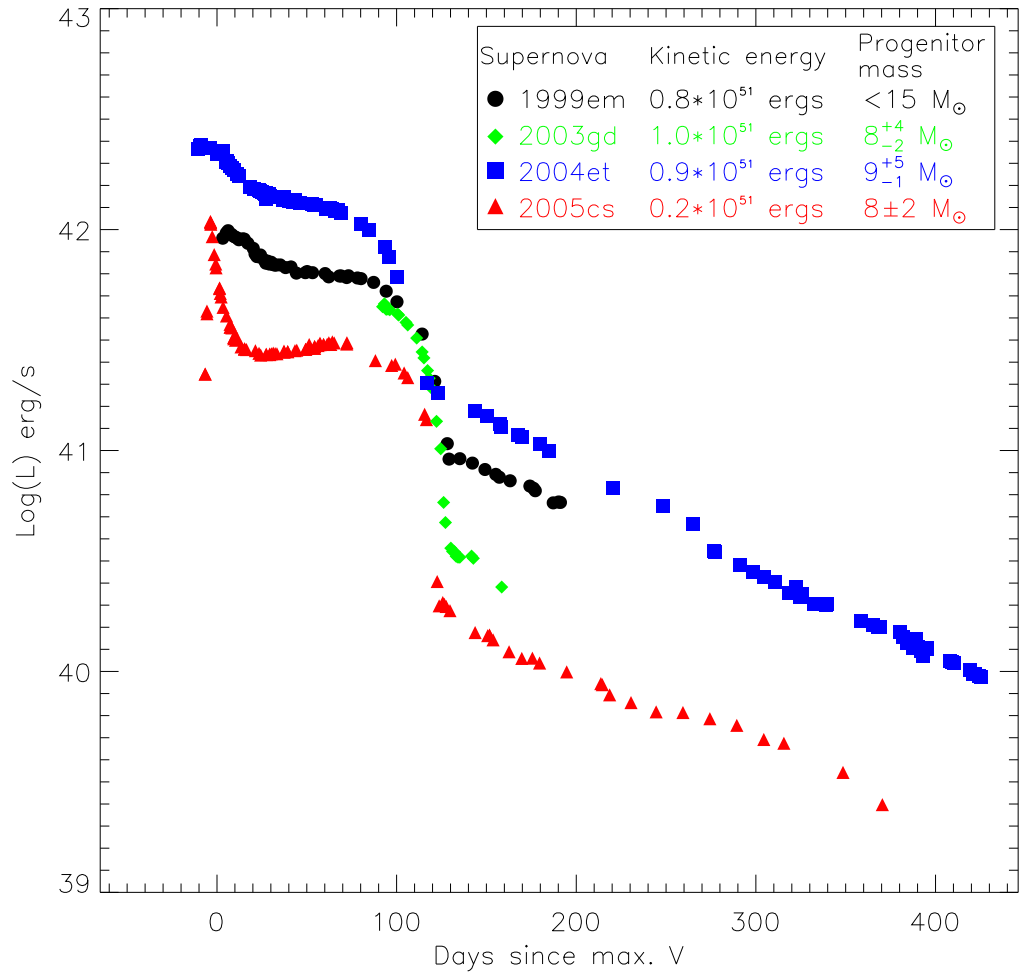,width=15cm,angle=0}}
\caption{ Bolometric lightcurves of II-P SNe. These four are likely to have had
similar progenitor stars and the progenitors of SN2003gd and SN2005cs appear to
be identical. There is a  large diversity in bolometric luminosity, kinetic energy and 
$^{56}$Ni mass from similar progenitors, hinting at intrinsic differences in the explosions. Data sources are 
SN1999em : \cite{2003MNRAS.338..939E}; 
SN2003gd : \cite{2005MNRAS.359..906H}; 
SN2005cs: \cite{2009MNRAS.394.2266P}
SN2004et :   \cite{2007MNRAS.381..280M}
}
\label{fig:II-P-BLCs}
\end{figure}

\newpage
\begin{figure}[ht]
\centerline{\psfig{file=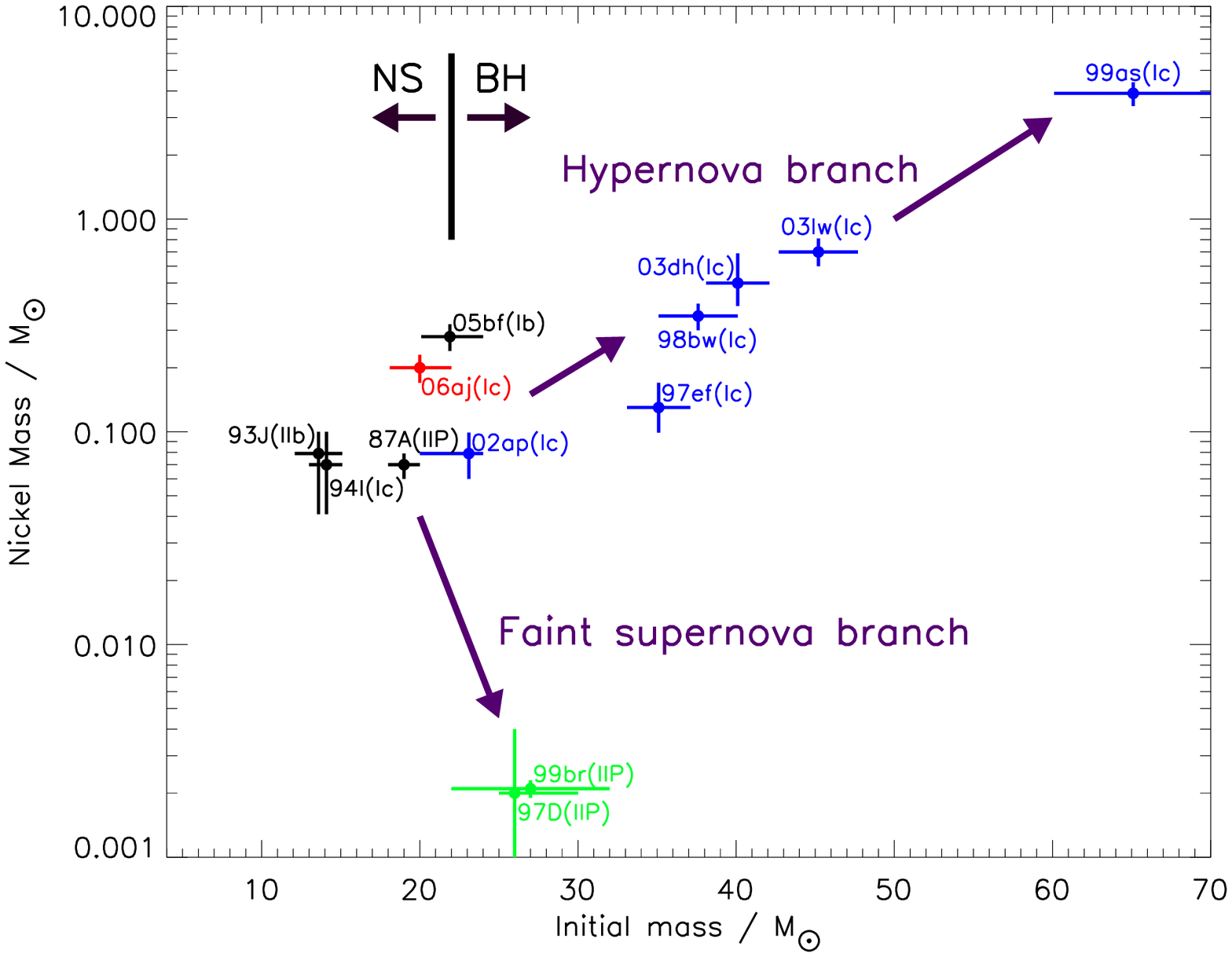,width=12cm,angle=0}}
\centerline{\psfig{file=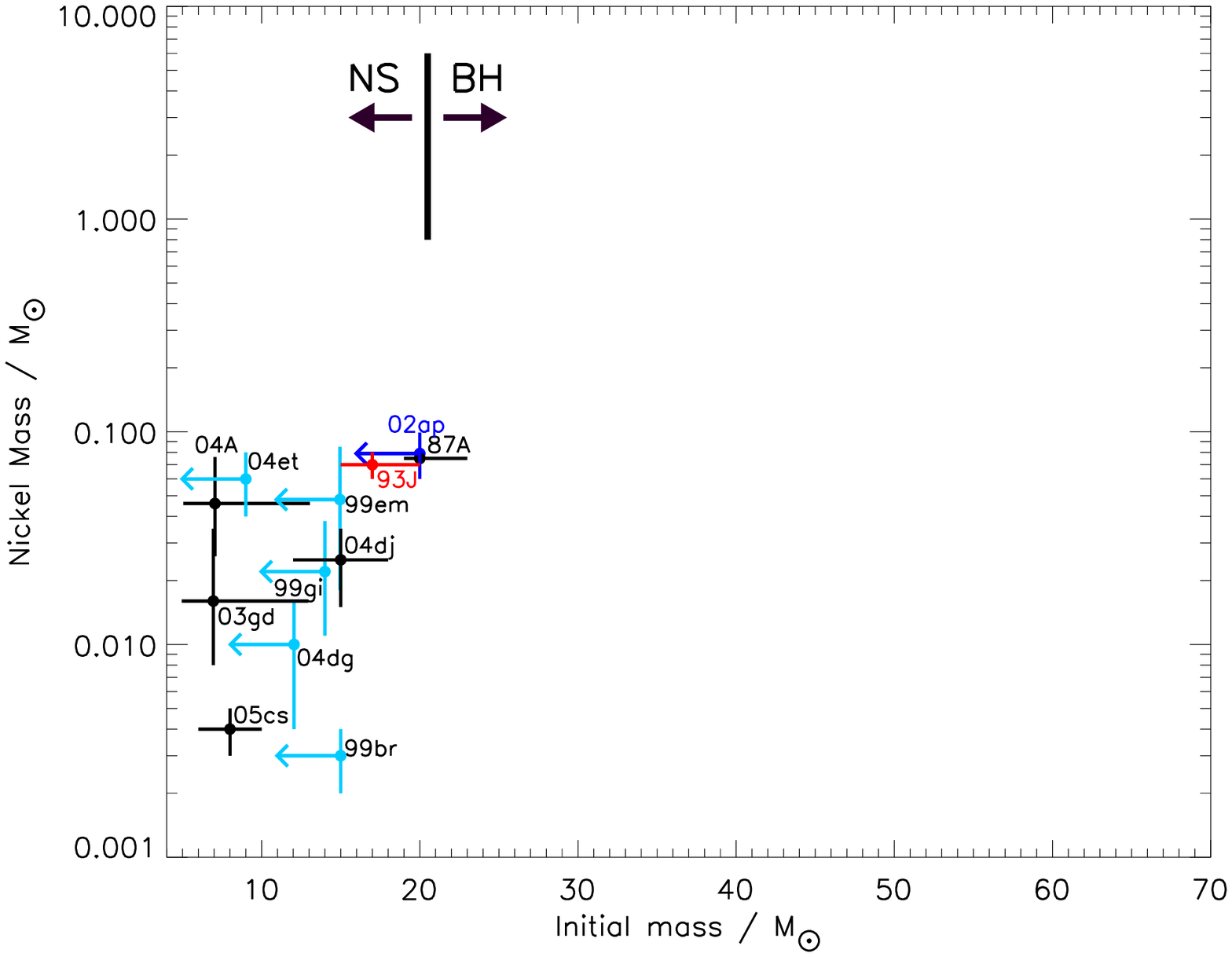,width=12cm,angle=0}}
\caption{\small $^{56}$Ni mass vs main-sequence initial mass with the upper
panel taken from 
\cite{2006NuPhA.777..424N} and the lower plot from 
\cite{SECM08}. The initial masses in this plot are 
estimated from the ejecta masses derived from lightcurve modelling. The lower
plot shows the $^{56}$Ni masses for nearby SNe for which there are reliable
restrictions on the progenitor masses from direct constraints.
}
\label{fig:Nimass}
\end{figure}

\begin{figure}[ht]
\centerline{\psfig{file=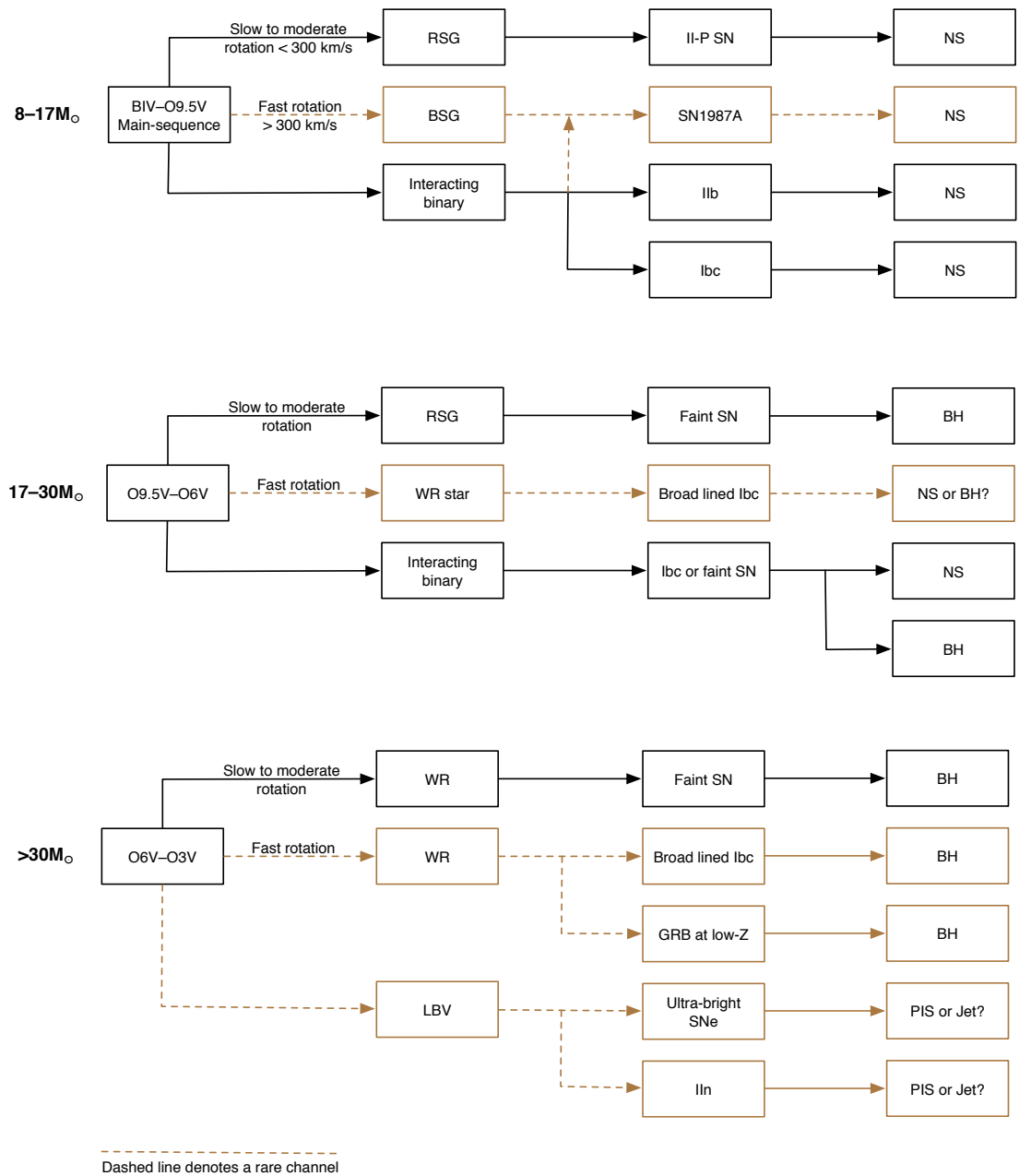,width=15cm,angle=0}}
\caption{A summary diagram of the likely evolutionary scenarios and end states
of massive stars, based on the observational evidence presented in this review.
The acronyms are neutron star (NS), black hole (BH), pair instability supernova (PIS).  
The probable rare channels of evolution are shown in light brown. The 
faint SNe are proposed and have not yet been detected. 
}

\label{fig:summary}
\end{figure}


\end{document}